\documentclass[11pt]{article}
 \usepackage{amsmath, xypic, graphicx}
\usepackage{stmaryrd,xspace,amssymb}
\usepackage{url}

\newcommand{\downset}{\ensuremath{\mathop{\downarrow\!}}}

\newcommand{\Q}{\mathbb{Q}}

\newcommand{\drie}{\vartriangleleft}

\newcommand{\beq}{\begin{equation}}
\newcommand{\eeq}{\end{equation}}
\newcommand{\bea}{\begin{eqnarray}}
\newcommand{\eea}{\end{eqnarray}} \newcommand{\nn}{\nonumber}

\newcommand{\Sets}{\mbox{\textbf{Sets}}}
\newcommand{\sr}{\stackrel}

\newcommand{\ca}{C*-algebra}

 \newcommand{\ovl}{\overline}
 \newcommand{\til}{\tilde}
\newcommand{\raw}{\rightarrow} 
\newcommand{\rac}{\rightarrowtail}

\newcommand{\law}{\leftarrow} \newcommand{\Raw}{\Rightarrow}

 \newcommand{\wed}{\wedge}

\newcommand{\x}{\times}

\newcommand{\inv}{^{-1}}

\newcommand{\er}{\eqref}
 
 \newcommand{\Gm}{\Gamma}
\newcommand{\dl}{\delta} \newcommand{\Dl}{\Delta}

\newcommand{\lm}{\lambda} 
 \newcommand{\sg}{\sigma}
\newcommand{\Sg}{\Sigma} \newcommand{\ta}{\tau} 
 \newcommand{\phv}{\varphi}
\newcommand{\ch}{\chi} \newcommand{\ps}{\psi} \newcommand{\Ps}{\Psi}
\newcommand{\om}{\omega} \newcommand{\Om}{\Omega}
\newcommand{\CA}{{\mathcal A}} 
\newcommand{\CC}{{\mathcal C}} 
 \newcommand{\CI}{{\mathcal I}}
   \newcommand{\CL}{{\mathcal L}}

\newcommand{\CO}{{\mathcal O}} \newcommand{\CP}{{\mathcal P}}
 
\newcommand{\CT}{{\mathcal T}} 
 
\newcommand{\C}{{\mathbb C}} 
 
\newcommand{\N}{{\mathbb N}} \newcommand{\R}{{\mathbb R}}
\newcommand{\T}{{\mathbb T}} 
\newcommand{\Hom}{\mbox{\rm Hom}}

\newcommand{\alg}[1]{\ensuremath{#1}}
\newcommand{\functor}[1]{\ensuremath{\underline{#1}}}

\newcommand{\id}[1]{\ensuremath{\mathrm{id}}}

\newcommand{\Sh}{\ensuremath{\mathrm{Sh}}}

\newcommand{\context}{\ensuremath{\mathcal{C}}}

\newcommand{\asstopos}{\ensuremath{\mathcal{T}}}
\newcommand{\interpretation}[1]{\ensuremath{\llbracket{#1}\rrbracket}}

\newcommand{\sa}{\ensuremath{_{\mathrm{sa}}}}

\newcommand{\field}[1]{\ensuremath{\mathbb{#1}}}

\newcommand{\uS}{\underline{\Sigma}}
\newcommand{\uA}{\underline{A}}

\renewcommand{\CA}{\mathcal{C}(A)}
\newcommand{\TA}{\mathcal{T}(A)}

\newcommand{\ulS}{\functor{\Sigma}}
\newcommand{\ulR}{\underline{\mathbb{R}}}
\newcommand{\ulA}{\underline{A}}
\renewcommand{\TA}{\asstopos(\alg{A})}
\renewcommand{\CA}{\context(\alg{A})}

{ \begin{trivlist}%
  \item[\hskip \labelsep {\bfseries #1}]%
}%
{ \end{trivlist}%
}

\topmargin = - 0.5 cm
\long\def\symbolfootnotemark[#1]{\begingroup%
\def\thefootnote{\fnsymbol{footnote}}\footnotemark[#1]\endgroup} 

\long\def\symbolfootnotetext[#1]#2{\begingroup%
\def\thefootnote{\fnsymbol{footnote}}\footnotetext[#1]{#2}\endgroup} 
\newcommand{\Mt}{\ensuremath{M_2(\C)}}
\newcommand{\uOm}{\underline{\Omega}}
\newcommand{\uF}{\underline{F}}
\newcommand{\OS}{\mathcal{O}(\underline{\Sigma})}
\newcommand{\ONS}{\mathcal{O}(\Sigma)}
\newcommand{\uSg}{\underline{\Sigma}}
\newcommand{\uL}{\underline{L}}
\newcommand{\ueen}{\underline{1}}

\begin{document}
\title{Intuitionistic quantum logic of an $n$-level system}
\author{
  Martijn Caspers\symbolfootnotemark[1]
  \and Chris Heunen\symbolfootnotemark[1] \symbolfootnotemark[2]
  \and Nicolaas P. Landsman\symbolfootnotemark[1]
  \and Bas Spitters\symbolfootnotemark[3]
}
\symbolfootnotetext[1]{
    Radboud Universiteit Nijmegen,
    Institute for Mathematics, Astrophysics, and Particle Physics,
 Heyendaalseweg 135, 6525 AJ    NIJMEGEN, THE NETHERLANDS. 
}
\symbolfootnotetext[2]{
    Radboud Universiteit Nijmegen,
    Institute for Computing and Information Sciences, Heyendaalseweg 135, 6525 AJ 
 NIJMEGEN, THE NETHERLANDS. 
}
\symbolfootnotetext[3]{
    Eindhoven University of Technology,
    Department of Mathematics and Computer Science, P.O. Box 513, 5600
    MB EINDHOVEN, THE NETHERLANDS. 
}\maketitle
\vspace*{-0.75cm}
\begin{center}{\it
Dedicated to Pekka Lahti, at his 60th birthday}
\end{center}
\smallskip
\begin{abstract}
A decade ago, Isham and Butterfield proposed a topos-theoretic approach to quantum mechanics, which meanwhile has been extended by D\"{o}ring and Isham so as to provide a new mathematical foundation for all of physics. Last year, 
three of the present authors redeveloped and refined these ideas by combining the C*-algebraic approach to quantum theory with the so-called internal language of topos theory (see arXiv:0709.4364). The goal of the present paper is to illustrate our abstract setup through the concrete example of the C*-algebra $M_n(\C)$ of complex $n\times n$ matrices. This leads to an explicit expression for the pointfree quantum phase space $\Sg_n$ and the associated logical structure and Gelfand transform of an $n$-level system. We also determine the pertinent non-probabilisitic state-proposition pairing (or valuation) and give a very natural topos-theoretic  reformulation of the Kochen--Specker Theorem.

In our approach, the {\it nondistributive} lattice $\CP(M_n(\C))$ of projections  in $M_n(\C)$ (which forms the basis of the  traditional quantum logic of Birkhoff and von Neumann) is replaced by a specific {\it distributive} lattice $\CO(\Sg_n)$ of functions from the poset $\CC(M_n(\C))$ of all unital commutative C*-subalgebras $C$ of $M_n(\C)$ to $\CP(M_n(\C))$. The lattice $\CO(\Sg_n)$  is essentially the (pointfree) topology of the quantum phase space $\Sg_n$, and as such defines a
 Heyting algebra. Each element of $\CO(\Sg_n)$ corresponds to a ``Bohrified'' proposition,  in the sense that  to each 
 classical context $C\in \CC(M_n(\C))$ it associates a yes-no question (i.e.\ an element of the Boolean lattice $\CP(C)$ of projections in $C$), rather than being a single projection as in standard quantum logic.
Distributivity is recovered at the expense of the law of the excluded middle ({\it Tertium Non Datur}), whose demise is in our opinion to be welcomed, not just
in intuitionistic logic in the spirit of Brouwer, but also in quantum logic  in the spirit of von Neumann.
\end{abstract}
\begin{center}\textbf{Motto}
\begin{quote}
`All departures from common language and ordinary logic are entirely avoided by reserving the word ``phenomenon'' solely for reference to unambiguously communicable information, in the account of which the word ``measurement'' is used in its plain meaning of standardized comparison.' (N. Bohr \cite{Bohrlogic})
\end{quote}
\end{center}

\newpage
\section{Introduction}
The main novelty of quantum mechanics, which also lies at the root of the difficulties in interpreting this theory, 
is the property that its truth attributions are ontologically (and not just epistemically) probabilistic. That is,
if $a\in\Dl$ denotes the proposition that an observable $a$ (represented by a self-adjoint operator on a Hilbert space $H$)
takes values in a (measurable) subset $\Dl\subset\R$, then even a \textit{pure} state $\ps$ (represented by a unit vector $\Ps$ in $H$) only gives a probabilistic truth attribution through the Born rule 
\beq \langle \ps,a\in \Dl\rangle=\| [a\in\Dl]\Ps\|^2.\label{Born}\eeq
Here the left-hand side denotes the probability that $a\in\Dl$ is true in the state $\ps$, and the expression $[a\in\Dl]$ in the right-hand side stands for the spectral projection defined by $a$ and $\Dl$ (often written as $E(\Dl)$ with $a$ understood).
In particular, unless $\Ps$ lies either in the image of $[a\in\Dl]$ (so that  the right-hand side equals one), or in the orthogonal complement thereof (in which case it is zero), one cannot say without running into contradictions whether or not the proposition $a\in\Dl$ is true. 

One of the aims of the topos-theoretic approach to quantum theory initiated by Isham and Butterfield \cite{BI2} (and subsequently extended by  D\"{o}ring and Isham so as to provide a new mathematical foundation for all of physics
\cite{DI}) is to define non-probabilistic
 truth attributions. Since one cannot just go back to classical physics, the price one has to pay for this is that such attributions
 do not take values in the set $\{0,1\}$ (identified with $\{\mathrm{false,true}\}$), but in some more general and abstract ``truth object'' $\Om$. Topos theory\footnote{This paper requires some familiarity with elementary category theory at the level of the first few chapters of \cite{ML}. The Appendix below contains sufficient  information on  topos theory to read this paper; as a general introduction to topos theory we recommend \cite{Gold}
for beginners and \cite{MLM} for readers already familiar with  category theory.  For the moment, it suffices to know that
a topos is a category in which most mathematical reasoning that one is familiar with in the category \Sets\ of sets and functions continues to make sense, with the exception that all proofs now have to be constructive (in the sense that in principle neither the law of excluded middle nor the axiom of choice may be used).  
}  provides natural candidates for such objects (under the name of subobject classifiers), and therefore seems to provide an appropriate tool for the search for the non-probabilistic essence of quantum theory.
  
To explain our setup, let us go back to classical physics for a moment. Let $M$ be the phase space of some physical system, with topology $\CO(M)$ (i.e.\ the subset of the power set of $M$ consisting of all open sets in $M$). We represent observables
by continuous functions $a:M\raw\R$, and once again consider propositions of the form $a\in\Dl$, with $\Dl\subset\R$ open.
For a pure state $x\in M$, we say that $a\in\Dl$ is true in $x$ iff $a(x)\in\Dl$ or, equivalently, iff  $x\in  a\inv(\Dl)$. Otherwise, $a\in\Dl$ is false.
We now claim that the above truth attribution to the proposition $a\in\Dl$ by the  pure state $x$, which we call
$\langle x,a\in \Dl\rangle$,
 may be captured in categorical terms in the following way \cite{HLS}:
\beq
   \big(1\stackrel{\langle x,a\in \Dl\rangle}{\longrightarrow}\Om\big)\:\:=\:\:
   \big(1\sr{[a\in\Dl]}{\longrightarrow} \CO(M)
   \sr{\interpretation{\dl_x=1}}{\longrightarrow}\Om\big). \label{spp}
\eeq
First, all objects and arrows are taken to be in \Sets. For example,  $1$ denotes an arbitrary but fixed singleton set;
elements $a\in A$ are identified with arrows $a:1\raw A$. Thus $\langle x,a\in \Dl\rangle:1\raw\Om$ on the left-hand side
denotes a specific element of the 
 set $\Om=\{0,1\}$ (identified with $\{\mathrm{true, false}\}$, as above). Similarly,  $[a\in\Dl]:1\raw\CO(M)$ denotes an element of the set $\CO(M)$, i.e.\ an open subset of $M$, namely $a\inv(\Dl)$. Thus opens in $M$ may be identified with equivalence classes of propositions of the type $a\in\Dl$: the latter defines the open $[a\in\Dl]=a\inv(\Dl)$, and conversely $S\in
 \CO(M)$ corresponds to the equivalence class of all propositions  $a\in\Dl$ for which $a\inv(\Dl)=S$.
 
Subsequently,  let $\dl_x$ be the Dirac measure on $M$, defined by $\dl_x(U)=1$ if $x\in U$ and  $\dl_x(U)=0$ if not. For technical reasons, like all probability measures we regard $\dl_x$ as a map $\dl_x:\CO(M)\raw\R^+$ (rather than as a map defined on all measurable 
subsets of $M$; this entails no loss of generality if $M$ is locally compact and Hausdorff).
The notation $\interpretation{\dl_x=1}:\CO(M)\raw\Om$ stands for 
\beq \interpretation{\dl_x=1}=\ch_{\{U\in\CO(M)\mid x\in U\}}, \eeq 
i.e.
the function that maps $U\in\CO(M)$ to 1 whenever
$\dl_x(U)=1$, i.e.\ whenever $x\in U$, and  to 0 otherwise. Finally, \er{spp} means that the arrow on the left-hand side is defined as the composition of the two arrows on the right-hand side; we invite the reader to check that the ensuing element $\langle x,a\in \Dl\rangle\in\{0,1\}$ indeed equals 1 when $a(x)\in\Dl$ and equals zero otherwise. 
 
Returning to the opening of this Introduction, in von Neumann's  approach to quantum mechanics, phase space $M$ is replaced by  Hilbert space $H$, its topology $\CO(M)$ is replaced by the lattice $\CP(B(H))$ of projections on $H$, 
$x\in M$ becomes a unit vector $\Ps\in H$, the arrow $[a\in\Dl]:1\raw \CP(B(H))$ now stands for the spectral projection $E(\Dl)$ defined by $a$, and $\Om=\{0,1\}$ is turned into the interval $[0,1]$. Finally, the pertinent map
$\CP(B(H))\raw[0,1]$ is given by $p\mapsto \|p\Ps\|^2$. This recovers the Born rule \er{Born}, but the categorical formulation does not add anything to its understanding.
Instead, our slogan is: {\it truth is prior to probability}. Thus we will {\it first} construct a non-probabilistic state-proposition pairing,
which only in a {\it second}  step  will presumably reproduce the Born rule.\footnote{To accomplish this, the derivation of the Born rule in  \cite{NPLBorn} will have to be combined with the results of the present paper. The idea of a non-probabilistic state-proposition pairing (or valuation, as they called it) is due to Isham and Butterfield \cite{BI2}. }  

Our main ingredient
is a novel quantum analogue of classical phase space, or rather of its topology, given by the notion of a {\it frame} (see Appendix \ref{A3}). A frame is a generalized topology, so that we will denote the frame representing our quantum phase space by $\ONS$, even though there is no actual underlying space $\Sg$ whose topology it is; we will occasionally even use the symbol $\Sg$ itself and refer to it as a ``virtual'' (or ``pointfree'') space.\footnote{Similarly, in noncommutative geometry it has become quite customary to speak of ``noncommutative spaces'' without there actually being an underlying space in the classical sense.}
 The main reason why we prefer frames to lattices of projections on Hilbert space (or to more general orthomodular lattices) is that in their guise of Heyting algebras, frames offer an intuitionistic logic for quantum mechanics, which in being distributive is superior to the traditional quantum logic of Birkhoff and von Neumann \cite{BvN,Piron,Vara}. Namely, we feel the latter is:
 \begin{itemize}
\item  {\it too radical} in giving up distributivity (for one thing rendering it problematic to interpret the logical operations $\wed$ and $\vee$ as conjunction and disjunction, respectively);
\item   {\it not radical enough} in keeping the law of excluded middle (so that it falls victim to Schr\"{o}dinger's cat and the like).
\end{itemize}
 Indeed, the quantum logical structure carried by our quantum phase space $\Sg$ has exactly the opposite features:
 it is distributive but drops the law of excluded middle.\footnote{See \cite{Coecke} for an different intuitionistic perspective on quantum logic, which we will explore separately.}

In principle, our formalism is capable of coping with the most general quantum systems, described by some unital \ca\ $A$
of observables, but in what follows the reader may simply keep the case $A=M_n(\C)$ in mind, to which we will specialize in the main body of the paper.  The construction of $\ONS$ is based on a specific reading of Bohr's `doctrine of classical concepts' \cite{BohrEinstein}, which, roughly speaking, expresses that the quantum world can only be seen through classical glasses. We adopt a very specific mathematical reading of this philosophy, 
 namely that a noncommutative algebra of observables $A$ of some quantum system has to be studied through its (unital) commutative C*-subalgebras \cite{handbook}.
Hence we form the poset $\CA$ of all such subalgebras, partially ordered by (set-theoretic) inclusion (i.e.\ $C\leq D$ iff $C\subseteq D$). 
We then form the topos 
\beq
\TA=\Sets^{\CA}\label{defTA}
\eeq
of (covariant) functors from $\CA$, seen as a category, to \Sets\ (cf.\ Appendix \ref{A1}).  We will $\underline{\mathrm{underline}}$  objects in $\TA$. As a case in point, we define  the tautological functor
\beq \ulA:C\mapsto C, \label{intA}\eeq
which maps a point $C\in\CA$  to the corresponding commutative \ca\ $C\subset A$ (seen as a set);
for $C\subseteq D$ the map $\uA(C\leq D):\uA(C)\raw \uA(D)$ is just the inclusion $C\hookrightarrow D$. 
We call $\ulA$ the {\it Bohrification} of $A$. The key point is that $\uA$ is a commutative
\ca\ in $\TA$ under natural operations (see \cite{HLS}), so that according to the general theory of commutative \ca s in topoi \cite{BM} it has a Gelfand spectrum. The latter, called $\CO(\uSg)$,  is a frame  and hence at the same time a Heyting algebra in the topos $\TA$, carrying the (intuitionistic) logical structure of $A$. This structure is defined within the topos $\TA$ (i.e., ``internally''), and as such is hard to understand. Fortunately, $\OS$ admits a so-called ``external'' description through an associated frame (and hence Heyting algebra) in \Sets, called $\ONS$, which in many ways behaves like the topology of an underlying 
``quantum phase space'' $\Sg$ of the system. Its explicit description \er{bohr} is one of the central results of this paper. 
Even if $A$ is a von Neumann algebra (which is the case in our running example $A=M_n(\C)$), so that the projection lattice $\CP(A)$ is a quantum logic in the sense of Birkhoff and von Neumann \cite{Redei},
the intuitionistic (and hence distributive) quantum logic carried by $\ONS$ as a Heyting algebra is quite different from the (nondistributive)
quantum logic defined by $\CP(A)$. 

In the classical case of a commutative \ca\ $A$ in \Sets, the Riesz representation theorem  yields a bijective correspondence between states on $A$ and probability measures $\mu$ on the Gelfand spectrum $\Sg$ of $A$ (i.e.\ the locally compact Hausdorff space for which $A\cong C(\Sg,\C)$). This generalizes: a state on the initial (possibly noncommutative) \ca\ $A$ in \Sets\ defines  a
probability measure $\mu$ on the spectrum $\uS$ of the Bohrification $\uA$ of $A$ \cite{HLS}. To make this analogy technically correct, though, one has to redefine the notion of a measure on a topological space as a map whose domain is  the collection of open sets (rather than Borel sets), and whose image is the set of positive lower reals $\R^+_l$ (rather than Dedekind reals) \cite{J2}. This redefinition is obvious in one direction:\footnote{See \cite{CS2} for the general theory. It would be more correct to speak of a valuation rather than a measure.} a measure $\mu$ in the usual sense may simply be restricted to the open sets, and some value $q=\mu(U)$ defines the lower real 
$\downset q=\{r\in\Q\mid r<q\}\in\R_l$.

Consequently, each state $\ps$ on $A$ defines a probability ``measure'' $\mu$
on $\uSg$ in the technical sense of an arrow 
\beq
\CO(\uSg)\stackrel{\mu}{\raw}\ulR^+_l \label{arrow1}
\eeq
 in $\TA$, where $\ulR^+_l$ denotes the positive  lower reals {\it in the topos} $\TA$ (cf.\ Appendix \ref{A3}). One has an obvious arrow
$\ueen \stackrel{\downset 1}{\raw}\ulR^+_l$ (where $\ueen$ is the terminal object in $\TA$, see Appendix \ref{A2}),
 and hence a composite arrow \beq\big(\CO(\uSg)\stackrel{1}{\raw}\ulR^+_l\big)=\big(
\CO(\uSg)\sr{\exists !}{\longrightarrow}
\ueen\stackrel{\downset 1}{\raw}\ulR_l^+\big).\label{arrow2}\eeq Now, whenever one has a pair of arrows
$X\stackrel{\sg}{\raw}Y$ and $X\stackrel{\ta}{\raw}Y$ in a topos $\CT$, one obtains an arrow
$X\stackrel{\interpretation{\sg=\ta}}{\longrightarrow}\Om$, where $\Om$ is the subobject classifier in $\CT$ \cite{MLM}. Applying this to \er{arrow1} and \er{arrow2},  we obtain  an arrow $\CO(\uSg)\stackrel{\interpretation{\mu= 1}}{\longrightarrow}\uOm$, which we compose with an arrow $\ueen\sr{S}{\raw}\OS$;
the latter plays the role of an open subset $S$
of $\uSg$ and defines an elementary propositions, exactly as in classical physics.

Our non-probabilistic state-proposition pairing $\langle \ps,S\rangle$, then,  is obtained by combining these arrows as in
\beq \big(\ueen\stackrel{\langle \ps,S\rangle}{\longrightarrow}\uOm\big)
\:=\: \big(\ueen \stackrel{S}{\raw} \CO(\uSg) \stackrel{\interpretation{\mu= 1}}{\longrightarrow}\uOm\big). \label{uom}
\eeq
This is our {\it Umdeutung} \cite{Heis} or ``quantum-mechanical reinterpretation'' of the corresponding classical expression 
\beq
   \big(1\stackrel{\langle x,S\rangle}{\longrightarrow}\Om\big)\:\:=\:\:
   \big(1\sr{S}{\longrightarrow} \CO(M)
   \sr{\interpretation{\dl_x=1}}{\longrightarrow}\Om\big), \label{sppbis}
\eeq
which is obtained from \er{spp} by replacing the open subset $[a\in\Dl]=a\inv(\Dl)$ by an arbitrary open $S\in \CO(M)$.\footnote{There is also a rather complicated ``quantum-mechanical reinterpretation'' of the classical expression $a\inv(\Dl)$ as an arrow $[a\in\Dl]: \ueen\raw\OS$, namely the `Daseinization' map originally proposed by D\"{o}ring and Isham \cite{DI} as redefined in 
 \cite{HLS}.} 

\bigskip

\noindent The plan of this paper is as follows.
 Following some preliminary calculations in Section \ref{s2}, we explicitly compute the Gelfand spectrum of $\underline{M_n(\C)}$  in Section \ref{s3}.
By definition, this  yields our quantum phase space, both in its internal description $\OS$ and in its external description $\ONS$. The latter
 carries the intuitionistic logical structure of an $n$-level quantum system, which is explicitly described in Section \ref{s4}. In fact, this section can be understood without any knowledge of topos theory, based as it is on the concrete expression  \er{bohr} for $\ONS$. 
 Section \ref{s6} is a rather technical intermezzo, in which we explicitly compute the Gelfand transform of the Bohrification $\uA$. This material is instructive in itself, but it is also necessary preparation for 
Section \ref{s5}, which elaborates our formulation  of the Kochen--Specker Theorem \cite{HLS}
in the spirit of Isham and Butterfield \cite{BI2}, i.e.\ as claiming the nonexistence of points of a certain ``space''. Our space, however appears to us to be much more natural than the one in  \cite{BI2}. Furthermore, our reformulation suggests a new proof of the Kochen--Specker Theorem
on the basis of intrinsic tools from topos theory (as opposed to previous topos-theoretic reformulations of this theorem \cite{BI2,HLS} whose proofs relied on the original theorem \cite{KS}). In
Section \ref{s7} we compute the non-probabilistic state-proposition pairing explained above, leading to the explicit formula \er{result}. Section \ref{s8} gives a concrete parametrization of the poset $\CA$ of unital commutative C*-subalgebras of $A$. 
Finally, the Appendix in three parts gives some background on sheaf theory, topos theory, and Heyting algebras and frames, respectively.
\subsection*{Acknowledgement}
The authors are indebted to Lotte Hollands and Steve Vickers for help with Section \ref{s8}, and to Henrik Zinkernagel
for providing the Bohr quote on the title page.  Ieke Moerdijk and Pedro Resende independently suggested us to take a look at the fundamental paper of Joyal and Tierney \cite{JT}, which indeed greatly facilitated a number of our computations. Finally, we benefited from a detailed referee report, which led to a number of clarifications. 
\section{A fresh look at the spectrum of $\C^n$}\setcounter{equation}{0}\label{s2}
We assume all \ca s to have a unit. 
The Gelfand spectrum $\Sg_A$ of a commutative \ca\ $A$ (denoted by $\Sg$ whenever it is clear which $A$ is meant) is usually defined as the set of nonzero multiplicative linear functionals on $A$, and coincides with the pure state space of $A$. It is necessary for our purposes to deal with the (Gelfand) topology $\CO(\Sg_A)$ directly, equipped with its natural structure as a frame (cf.\ Appendix \ref{A3}). This frame can be constructed in a way that generalizes to  commutative \ca s in topoi. This was first shown by Banaschewski and Mulvey (see \cite{BM} and references to earlier work therein), but for our present computational purposes it is more convenient to follow the reformulation of the constructive theory of the Gelfand spectrum in \cite{Coq} (see also \cite{CS0,CS1,HLS}). In what follows, $A\sa=\{a\in A\mid a^*=a\}$ denotes the self-adjoint part of a \ca\ $A$, and $A^+=\{a\in A\sa\mid a\geq 0\}$ is its positive cone. 
\begin{enumerate}
\item The positive cone $A^+$ of a commutative \ca\ $A$ is a distributive lattice. Define an equivalence relation $\sim$
on $A^+$ by putting $a\sim b$ whenever
there are integers $n,m\in\N$ such that $a\leq nb$ and $b\leq ma$. The quotient
\beq 
L_A=A^+/\sim \label{quo}
\eeq is again a distributive lattice. Decomposing $a\in A\sa$ as $a=a^+-a^-$ (with $a^{\pm}\in A^+$) in the usual way, there is a surjective lattice homomorphism
 $A\sa\raw L_A$ given by $a\mapsto [a^+]$ (which restricts to the canonical
 projection $A^+\raw L_A$ for $a\in A^+$).
\item For $a\in A^+$ with image $[a]\in L_A$ and $U\in \mathbb{P}(L_A)$ (i.e.\ the power set of $L_A$), we say that $[a]\drie U$ iff  for all  
 $q>0$ there exists a  finite subset $U_0\subseteq U$ such that $ [(a-q\cdot 1)^+]\leq\bigvee U_0$.
\item The frame $\CO(\Sg_A)$ is given by 
\beq \CO(\Sg_A)=\{U\in DL_A\mid x\drie U \Raw x\in U\}, \label{UDL} \eeq
where $DL_A$ is the poset of all lower sets in $L_A$, ordered by  set-theoretic  inclusion.\footnote{A {\it lower} set in a poset $P$
is a subset $L\subseteq P$ such that $x\in L$ and $y\leq x$ implies $y\in L$.}
\end{enumerate}

This procedure simplifies when $A$ is finite-dimensional, in which case $L_A$ is a finite lattice. In that case, since $[(a-q)^+]=[a]$ for small enough $q$, one simply has $x\drie U$ iff $x\leq \bigvee U$, and the condition $x\drie U \Raw x\in U$
in \er{UDL} holds iff $U$ is a (principal) down set, i.e.\ $U=\downarrow\!x$ for some $x\in L_A$ (not the same $x$ as the placeholder $x$ in \er{UDL}). Hence for  finite-dimensional $A$ we have
\beq \CO(\Sg_A)=\{\downarrow\!x\mid x\in L_A\}.\label{fds}\eeq

For $A=\C^n$, step 1 yields $A^+=(\R^n)^+$. One has $(r_1,\ldots, r_n)\sim (s_1,\ldots, s_n)$ just in case that $r_i=0$ iff $s_i=0$ for all $i=1,\ldots n$.
Hence each equivalence class under $\sim$ has a unique representative of the form $[k_1,\ldots, k_n]$ with $k_i=0$ or $k_i=1$; the preimages of such an element of $L_A$ in $A^+$ under the natural projection $A^+\raw A^+/\sim$ are the  diagonal matrices whose $i$'th entry is zero if $k_i=0$ and any nonzero positive number if $k_i=1$. The partial order in $L_A$ is pointwise, i.e.\ $[k_1,\ldots, k_n]\leq [l_1,\ldots, l_n]$ iff $k_i\leq l_i$ for all $i$. Hence $L_{\C^n}$ is isomorphic
as a distributive lattice to the lattice $\CP(\C^n)$ of projections in $\C^n$, i.e.\ the
lattice of those projections in $M_n(\C)$ that are diagonal matrices: under this isomorphism $[k_1,\ldots, k_n]$ corresponds
 to the  diagonal matrix  $\mathrm{diag}(k_1,\ldots, k_n)$. If we equip  $\CP(\C^n)$ with the usual partial ordering
of projections on the Hilbert space $\C^n$, viz.\ $p\leq q$ whenever $p\,\C^n\subseteq q\,\C^n$ (which coincides with their ordering as element of positive cone of the \ca\ $M_n(\C)$), then this is even a lattice isomorphism.

Consequently, $\CO(\Sg_{\C^n})$ consists of all sets of the form $\downset\, p$, $p\in \CP(\C^n)$,
 partially ordered by inclusion. 
 Of course, this means that 
 \beq
 \CO(\Sg_{\C^n})\cong\CP(\C^n),\label{2.4}\eeq
  under the further identification of $\downset\,p\in \mathbb{P}(\CP(\C^n))$ with $p\in\CP(\C^n)$.
  This starts out just as an
isomorphism of posets, and turns out to be one of frames (which in the case at hand happen to be Boolean).
To draw the connection with the usual spectrum $\hat{\C}^n=\{1,2,\ldots, n\}$ of $\C^n$, we note that
the right-hand side of \er{2.4}  is isomorphic to the discrete topology  $\CO(\hat{\C}^n)=\mathbb{P}(\hat{\C}^n)$  of $\hat{\C}^n$
under the isomorphism (of lattices and even of frames)
\begin{eqnarray}
 \CP(\C^n)& \sr{\cong}{\raw} & \CO(\hat{\C}^n); \nn \\
 \mathrm{diag}(k_1,\ldots, k_n)&\mapsto& \{i\in \{1,2,\ldots, n\} \mid k_i=1\}. \label{GS}
\end{eqnarray}

We now describe the Gelfand transform, which in general is given by 
\begin{eqnarray}
A&\sr{\cong}{\raw}& C(\Sg,\C); \nn \\
a&\mapsto&\hat{a};\nn \\
 \hat{a}(\om)&=&\om(a).
\end{eqnarray}

Let $a=(a_1,\ldots, a_n)\in \C^n\sa=\R^n$.  With $\Sg$ realized as $\hat{\C}^n$, this just reads 
 \beq \hat{a}(i)=a_i,\eeq for $\hat{a}:\hat{\C}^n\raw\C$.  The induced frame map is given by
 \begin{eqnarray}
 \hat{a}\inv:\CO(\C)&\raw & \CO(\hat{\C}^n);\nn \\
  U&\mapsto&  \{i\in \{1,2,\ldots, n\} \mid a_i\in U\}.
\end{eqnarray}
By \er{GS}, this is  equivalent to
  \begin{eqnarray}
 \hat{a}\inv:\CO(\C)&\raw &  \CP(\C^n)   ;\nn \\
  U&\mapsto&  \mathrm{diag}(\ch_U(a_1),\ldots,\ch_U(a_n)).\label{GT0}
\end{eqnarray}
For $a=a^*$ we may regard the Gelfand transform as an isomorphism $A\sa\sr{\cong}{\raw} C(\Sg,\R)$, and \er{GT0} may be rewritten as 
  \begin{eqnarray}
 \hat{a}\inv:\CO(\R)&\raw & \CP(\C^n) ;\nn \\
  U&\mapsto& [a\in U], \label{GTSP}
\end{eqnarray}
  where $U\in\CO(\R)$, and the right-hand side denotes the spectral projection $\ch_U(a)$ defined by the self-adjoint operator $a$ on the Hilbert space $\C^n$. 
  
 Finally, we write the usual correspondence between states $\ps$ on a commutative \ca\ $A$ and probability measures $\mu_{\ps}\equiv\mu$ on its spectrum $\Sg$ in a way that can be generalized to topos theory. In the usual setting, one may define the value of $\mu$ on each open $E\in\CO(\Sg)$ by means of
 \beq
 \mu(E)=\sup\{\ps(a)\mid 0\leq a\leq 1, \mathrm{supp}(\hat{a})\subset E\},\eeq
 where instead of $0\leq a\leq 1$ we could just as well write $0\leq \hat{a}\leq 1$.
Using \er{2.4}, for  $A=\C^n$ this implies that $\mu:  \CP(\C^n)\raw \R^+$ is given by
 \beq \mu(p)=\ps(p). \label{mup}\eeq
\section{Spectrum of the Bohrification of $M_n(\C)$}\setcounter{equation}{0}\label{s3}
We now use the insights of the previous section to compute the spectrum $\uSg_{\underline{A}}$ --- more precisely, its associated frame 
 $\CO(\uSg_{\uA})$ ---
 of the Bohrification $\ulA$ of $A=M_n(\C)$ in the  topos $\TA$. This may be done by implementing the three-step program of the previous section, giving the appropriate topos-theoretical meanings to the various constructions involved.\footnote{Technically, this means that one has to use the internal or Mitchell-B\'{e}nabou language of the topos.}  
For simplicity we write $\uL$ for the lattice $L_{\ulA}$  in $\TA$;
similarly, $\uSg$ stands for $\uSg_{\underline{A}}$. 

 To begin with,  for arbitrary $A$ the lattice $\uL$ can  be computed ``locally'', in the sense that $\uL(C)=L_C$ \cite{HLS}, so that by \er{quo} one has $\uL(C)=C^+/\sim$. Let $\CP(C)$ be the (Boolean)  lattice of projections in $C$, and 
 consider the functor $C\mapsto\CP(C)$, where 
 the arrow $C\subseteq D$ in $\CA$ induces the inclusion map $\CP(C)\hookrightarrow \CP(D)$. It follows (cf.\  the preceding section) that we may identify ${\uL}(C)$ with $\CP(C)$ and hence we may identify the functor
   ${\uL}$ with the functor $\CP$.  We will make this identification in what follows.
 
Second,  whereas in \Sets\ \er{fds} makes $\CO(\Sg)$ a subset of $L$, in the topos $\TA$ the frame $\OS$ is a subobject $\CO(\uSg)\rac \uOm^{\uL}$, where $\uOm$ is the subobject classifier in $\TA$. It then follows from \er{Yoneda} that $\OS(C)$ is a certain subset of $\mathrm{Sub}(\CP_{\uparrow C})$, the set of subfunctors of the functor $\CP:\CA\raw\Sets$ restricted to $\uparrow\! C\subset\CA$. To explain which subset, define
\begin{eqnarray}
\mathrm{Sub}_d(\CP_{\uparrow C})&=&\{\til{S}\in \mathrm{Sub}(\CP_{\uparrow C})\mid \forall \, D\supseteq C\;
\exists\,  x_D\in \CP(D): \til{S}(D)=\downarrow\! x_D\}; \label{bij} 
\end{eqnarray}
In other words, $\mathrm{Sub}_d(\CP_{\uparrow C})$ consists 
of those  subfunctors $S$ of $\CP_{\uparrow C}$ that are locally (principal) down-sets.
It then follows from \er{fds} and the local interpretation of the relation $\drie$ in $\TA$ \cite{HLS} that the subobject  $\CO(\uSg)\rac \uOm^{\uL}$ in $\TA$ is the functor \beq \CO(\uSg)(C)=\mathrm{Sub}_d(\CP_{\uparrow C});
 \label{SgC}\eeq
 the map $$\OS(C\leq D): \CO(\uSg)(C)\raw \CO(\uSg)(D),$$ defined
whenever $C\subseteq D$, is inherited from $\uOm^{\uL}$ (of which $\OS$ is a subobject), and hence is
just given by restricting an element  of $\CO(\uSg)(C)$
 to $\uparrow\! D$.
 
 Writing
 \beq \mathrm{Sub}_d(\CP)=\{\til{S}\in \mathrm{Sub}(\CP)\mid \forall \, D\in\CA\;
\exists\,  x_D\in \CP(D): \til{S}(D)=\downarrow\! x_D\} ,\label{openS} \eeq
it is convenient to embed
 $\mathrm{Sub}_d(\CP_{\uparrow C})\subseteq \mathrm{Sub}_d(\CP)$ by requiring elements of the left-hand side
 to vanish whenever $D$ does not contain $C$. We also note that
 if $\til{S}$ is to be a subfunctor of $\CP_{\uparrow C}$, one must have $\til{S}(D)\subseteq \til{S}(E)$ whenever $D\subseteq E$, and that
$\downarrow\! x_D\subseteq \downarrow\! x_E$ iff $x_D\leq x_E$ in $\CP(E)$.
Thus
one may simply describe elements  of  $\CO(\uSg)(C)$ 
via maps $S:\mathcal{C}(A) \raw \CP(A)$ such that:
 \begin{enumerate}
 \item  ${S}(D)\in \CP(D)$;
\item  ${S}(D)=0$ whenever $D\notin\uparrow\! C$ (i.e.\ whenever $C\nsubseteq D$);
\item  ${S}(D)\leq {S}(E)$ whenever $C\subseteq D\subseteq E$.
\end{enumerate}
The corresponding element $\til{S}$ of  $\CO(\uSg)(C)$  is then given by
 \beq \til{S}(D)=\downset S(D),\label{down}\eeq
 seen as a subset of $\CP(D)$. Hence it is convenient to introduce the notation
 \begin{eqnarray}
\CO(\Sg)&=&\{S:\mathcal{C}(A) \raw \CP(A)\mid {S}(D)\in \CP(D),\, 
{S}(D)\leq {S}(E)\:\mbox{if}\: D\subseteq E\}; \label{bohr} \\
 \CO(\Sg)_{\uparrow C}&=&\{S:\,  \uparrow\! C\raw \CP(A)\mid {S}(D)\in \CP(D),\, 
{S}(D)\leq {S}(E)\:\mbox{if}\: D\subseteq E\}, \label{bohrC} 
\end{eqnarray}
in terms of which we have isomorphisms
\begin{eqnarray}
\CO(\uSg)(\C\cdot 1)&\cong& \ONS \label{keyiso};\\
\CO(\uSg)(C)&\cong&   \CO(\Sg)_{\uparrow C}\label{keyisoC}
\end{eqnarray}
 of posets and even of frames, provided we define the partial order on $\ONS$ 
 and $ \CO(\Sg)_{\uparrow C}$ pointwise with respect to
the usual ordering of projections, i.e.
\beq 
  S\leq T\: \Leftrightarrow \: S(D)\leq T(D) \mbox{ for all } D\in\CA.\label{posg}\eeq 
Although \er{bohr} and \er{keyiso} are special cases of \er{bohrC} and \er{keyisoC}, respectively
(namely for $C=\C\cdot 1$), we have singled them out for two reasons:
\begin{enumerate}
\item $\ONS$ plays a special role because of the isomorphism
   \beq \Hom(1, \CO(\uSg))\cong\CO(\uSg)(\C\cdot 1),
  \label{18} \eeq
 according to which each element $S$ in \er{bohr} may be seen as an ``open'' $1\stackrel{S}\raw  \CO(\uSg)$. 
 \item $\ONS$ is the key to the ``external'' description of $\OS$ (see Appendix \ref{A3}).
\end{enumerate}
Namely, equipping the poset 
 $\CA$ with the Alexandrov topology,\footnote{\label{AT} Let $P$ be a poset.
 The open subsets of $P$ in the Alexandrov topology are the upper sets, i.e.\ those $U\subseteq P$ for which $x\in U$ and $x\leq y$ implies $y\in U$.
Basic examples of such opens  are up-sets $U=\mathop{\uparrow}\!  x=\{y\in P\mid x\leq
y\}$, which form a basis of the  Alexandrov topology. In fact, ${\uparrow} x$ is the smallest open containing $x$.} this description is given by the frame map
\beq
 \pi_{\Sg}^*:\CO(\CA)\raw\ONS, \label{JTmap}
 \eeq
given  on the basic opens $\uparrow\! D\in \CO(\CA)$  by
\begin{eqnarray}
 \pi_{\Sg}^*(\uparrow\! D)=\ch_{\uparrow D}&:&  E\mapsto 1\:\: (E\supseteq D);\nn \\
& & E\mapsto 0\:\: (E\nsupseteq D).\label{defpi}
\end{eqnarray}
The  external description of $\OS$ will be put to good use in Sections \ref{s6} and \ref{s5}.

To close this section,  let us make two clarifying remarks.
\begin{enumerate}
\item  The key identification \er{bohr} eventually relies on $C\cong \C^k$ for some $k$, and on $\C^k$ having enough projections as a C*-algebra (hence a similar result holds for arbitrary von Neumann algebras \cite{HLSSyn}). 
\item The  D\"{o}ring--Isham `Daseinization' map \cite{DI,HLS} leads to a specific association of some $S\in
\CO(\Sg)$ with a quantum-mechanical proposition $a\in\Delta$ in the usual (von Neumann)
sense; cf.\ the Introduction. Thus we see that whereas in the traditional approach to quantum logic a {\it single} projection $p\in\CP(A)$ defines a proposition, in our `Bohrian' approach  a proposition $S$ consists of a {\it family} of projections $S(C)$, one  for each classical context $C$: this is our version of complementarity.\footnote{The connection between Bohr's notion of complementarity and the actual formalism of quantum mechanics is beautifully described
in two papers by our dedicatee \cite{Lahti1,Lahti2}.} See, however, \er{ronnie}  for a map $\CP(A)\raw\CO(\Sg)$ relating traditional and intuitionistic quantum logic.
\end{enumerate}
\section{Intuitionistic quantum logic of $M_n(\C)$}\setcounter{equation}{0}\label{s4}
Thus our claim is that $\ONS$ in \er{bohr} with $A=M_n(\C)$ 
describes the correct quantum logic of an $n$-level system. We will now determine the  Heyting algebra structure of $\ONS$. 

First,  the top and bottom elements of $\ONS$  are given by
\begin{eqnarray}
\top(C)&=&1\: \text{ for all }  C;\label{topS}\\
\bot(C)&=&0\: \text{ for all }  C. \label{botS}
\end{eqnarray}
The logical operations on  $\ONS$ may be computed from the partial order; cf.\ \cite[\S I.8]{MLM}. First, we obtain
\begin{eqnarray}
(S \wed T)(C) &=& S(C) \wed T(C); \\
(S \vee T)(C) &=& S(C) \vee T(C)
\end{eqnarray}
for $\inf$ and $\sup$. The operation of (Heyting) implication \er{fHA} is given by the more interesting expression\footnote{Naively, one would expect the right-hand side of \er{HI} to be $S(C)^{\perp}\vee T(C)$, but that would not define an element of $\ONS$. The notation  $\bigwedge^{\CP(C)}$ in \er{HI} indicates that one takes a greatest lower bound over all $S(D)^{\perp}\vee T(D)$ that is constrained to lie in $\CP(C)$. Analogously for the  lowest upper bound $\bigvee^{\CP(D)}$ in \er{notnot}.}
\beq 
(S\rightarrow T)(C)=\bigvee\{p\in \CP(C)\mid p\leq  S(D)^{\perp}\vee T(D)\:\forall\, D\supseteq C\}  \equiv\bigwedge^{\CP(C)}_{D\supseteq C}S(D)^{\perp}\vee T(D);
\label{HI}
\eeq recall that each ${S}$ is some projection in $D$, so that ${S}(D)^{\perp}=1-{S}(D)$.
 The derived operation of
negation \er{defneg} is therefore equal to 
\beq
(\neg S)(C)=\bigwedge^{\CP(C)}_{D\supseteq C}S(D)^{\perp}, \label{Hneg}
 \eeq
which yields
\begin{equation}
(\neg\neg {S})(C) = \bigwedge^{\CP(C)}_{D\supseteq C}\bigvee^{\CP(D)}_{E\supseteq D}{S}(E).
\label{notnot}
\end{equation}
In general, this is by no means equal to ${S}(C)$, so that our new quantum logic is indeed intuitionistic (as general topos theory suggests).
The failure of the law of excluded middle may be illustrated by the following example
for $A=M_3(\mathbb{C})$. In that case, one has (see Section \ref{s8})
\beq  \mathcal{C}(A) = \mathbb{C} \cdot 1\cup  \{U \cdot D_2 \cdot U^*\mid U \in SU(3) \} \cup
 \{U \cdot D_3 \cdot U^*\mid U \in SU(3) \},\eeq
 with
\begin{eqnarray}
   D_2   & =&  \{\mathrm{diag}(a,a,b) \mid a,b \in
  \mathbb{C}\}; \nn \\
     D_3   & =&  \{\mathrm{diag}(a,b,c) \mid a,b,c \in
  \mathbb{C}\}.
\end{eqnarray}
Now  define
${S} : \mathcal{C}(A) \to \mathcal{P}(A)$ by 
\begin{align}
    {S}(\mathbb{C} \cdot 1) 
  & = 0; \nn \\
    {S}(U \cdot D_2 \cdot U^*) 
  & = U\cdot \mathrm{diag}(1,1,0) \cdot U^*;\nn \\
  {S}(U \cdot D_3 \cdot U^*) 
  & = 1.
\end{align}
This indeed defines an element of $\ONS$; see \er{bohr}.
Then \er{notnot} yields
\beq
      (\neg\neg{S})(\C\cdot 1)=1,\eeq
 which is clearly different from  $S(\C\cdot 1)=0$.
 
 Finally, we briefly discuss an interesting map $\CP(A)\raw\CO(\Sg)$, $p\mapsto S_p$, given by\footnote{This map was independently proposed by our M.Sc.\ student Ronnie Hermens and by the referee.}
\begin{eqnarray}
S_p(C) &=& p \mbox{ if } p\in C;\nn \\
&=& 0 \mbox{ if } p\notin C. \label{ronnie}
\end{eqnarray}
This map may be seen as an extremely crude analogue of the   D\"{o}ring--Isham `Daseinization' map $\dl$ \cite{DI,HLS}, in which the approximant $\dl(p)_C$ to $p$
at $C$ is simply taken to be zero if $C$ does not contain $p$. However, this map fails to
preserve both $\vee$ and $\wed$ and is, in general, not even a lattice homomorphism when restricted to some Boolean part $\CP(C)\subset \CP(A)$.
 \section{Gelfand transform}\setcounter{equation}{0}\label{s6}
We now compute the Gelfand transform for $\uA$. For a general commutative \ca\ $\uA$ in a topos $\CT$  with spectrum $\uSg$ 
(which, we repeat, is really given by the corresponding frame $\OS$)
one has \cite{BM}
\beq
\uA\sa\cong C(\uSg,\ulR)\equiv \mathrm{Frm}(\CO(\ulR),\CO(\uSg)),\label{GTT}\eeq
where the right-hand side is the {\it definition} of the middle term (which is just a convenient notation). 
This reduces to the usual Gelfand transform in the topos \Sets, where $C(\Sg,\R)$ happens to have its usual meaning of continuous functions of $\Sg$ to $\R$.\footnote{\label{fnsober}
This is because in \Sets\ both $\uSg\equiv\Sg$ and $\ulR\equiv\R$ are so-called sober spaces  \cite[Def.\ IX.3.2]{MLM}, in which case $C(\Sg,\R)$ is isomorphic to $\mathrm{Frm}(\CO(\R),\CO(\Sg))$ 
 through $f\leftrightarrow f\inv$.}
To understand the general case, one has to distinguish between:
\begin{enumerate}
\item 
 the {\it set} $\Hom_{\mathrm{Frm}}(\CO(\ulR),\CO(\uSg))$
of internal frame maps from  the frame $\CO(\ulR)$ of (Dedekind) real numbers in $\CT$ to the frame $\OS$
(i.e., the set of those arrows from  $\CO(\ulR)$ to $\OS$ that happen to be frame maps as seen from within $\CT$);
\item
the {\it object} $\mathrm{Frm}(\CO(\ulR),\CO(\uSg))$ in $\CT$  defined as the subobject of the 
exponential $\OS^{\CO(\ulR)}$ consisting of (internal) frame maps  from  $\CO(\ulR)$ to $\OS$.
\end{enumerate}
The connection between 1.\ and 2.\ is given by the bijective correspondence \cite[p.\ 20]{MLM} between $C\raw B^A$ and $A\x C\raw B$; taking $C=1$ (the terminal object in $\CT$), we see that an {\it element}
of the set $\Hom(A,B)$ corresponds to an {\it arrow} $1\raw B^A$.

We now take $\CT=\TA$, in which case \er{NatC} yields
\beq
\mathrm{Frm}(\CO(\ulR),\CO(\uSg))(C)=\mathrm{Nat}_{\mathrm{Frm}}(\CO(\ulR)_{\uparrow C},\CO(\uSg)_{\uparrow C}),
\eeq
the set of all natural transformations between the functors $\CO(\ulR)$ and $\OS$, both restricted to
$\uparrow\! C\subset\CA$, that are frame maps. This set can be computed from  the external description of frames and frame  maps explained in Appendix \ref{A3}. As before, the poset 
 $\CA$ and its open subsets of the type $\uparrow\!C$ are equipped with the Alexandrov topology.

First, the frame $\CO(\ulR)_{\uparrow C}$ has external description
\beq
\pi\inv_{\R}:\,  \CO(\uparrow\!C)\raw \CO(\uparrow\!C\x\R),\eeq
 where $\pi_{\R}:\, \uparrow\!C\x\R\raw\uparrow\!C$ is projection on the first component. 
The special case $C=\C\cdot 1$ yields  the  external description
 of  $\CO(\ulR)$ itself, namely \beq
\pi\inv_{\R}:\,  \CO(\CA)\raw \CO(\CA\x\R),\eeq
 where this time (with some abuse of notation) the projection is  $\pi_{\R}:\, \CA\x\R\raw\CA$.
 
 Second, 
 the frame $\CO(\uSg)_{\uparrow C}$  has external description
 \beq  \pi_{\Sg}^*:\CO(\uparrow\!C)\raw\ONS_{\uparrow C},\eeq 
 given by \er{defpi} with the understanding that  $D\supseteq C$ (the
 special case  $C=\C\cdot 1$ then recovers  the external description \er{JTmap} of $\OS$ itself).
 
It follows that there is a bijective correspondence between frame maps $$\underline{\phv}_C^*:\CO(\ulR)_{\uparrow C}\raw\CO(\uSg)_{\uparrow C}$$ in $\TA$ and frame maps
\beq
\phv_C^*: \CO(\uparrow\! C\x\R)\raw \ONS_{\uparrow C}\label{phster}
\eeq
in \Sets\ that for any $D\supseteq C$ satisfy the condition
\beq \phv_C^*(\uparrow\! D\x\R)=\ch_{\uparrow D}.
\label{phcon}\eeq
Indeed,  such a map $\phv_C^*$ defines an element $\underline{\phv}_C^*$ of $\mathrm{Nat}(\CO(\ulR)_{\uparrow C},\CO(\uSg)_{\uparrow C})$  in the obvious way: for $D\in \uparrow\! C$, the components $\underline{\phv}_C^*(D):\CO(\ulR)(D)\raw\CO(\uSg)(D)$
of the natural transformation $\underline{\phv}_C^*$, i.e.\ \beq
\underline{\phv}_C^*(D): \CO(\uparrow\! D \times \R)\raw \ONS_{\uparrow D},
\label{phCD}\eeq  are simply given as the restriction of  $\phv_C^*$ to  $\CO(\uparrow\!\! D\x\R)\subset \CO(\uparrow\! C\x\R)$;
cf.\ \er{simplyput} in Appendix \ref{A3}. This is consistent, because \er{phcon} implies that 
 for any $U\in\CO(\R)$ and $C\subseteq D\subseteq E$ one has
\beq \phv_C^*(\uparrow\! E\x U)(F)\leq  \phv_C^*(\uparrow\! D\x \R)(F),\eeq
which  by \er{phcon} vanishes whenever $F\nsupseteq D$. Consequently, 
\beq
\phv_C^*(\uparrow\! E\x U)(F)=0\, \mbox{if}\,  F\nsupseteq D, \label{con2}\eeq
so that $\underline{\phv}_C^*(D)$ actually takes values in $ \ONS_{\uparrow D}$ (rather than in $\ONS_{\uparrow C}$, as might be expected).

Denoting the set of frame maps \er{phster} that satisfy \er{phcon}  by $\mathrm{Frm}'(\CO(\uparrow\! C\x\R),\ONS_{\uparrow C})$, we obtain a functor $\mathrm{Frm}'(\CO(\uparrow\! -\x\R),\ONS_{\uparrow -}):\CA\raw\Sets$, given by
\beq
C\mapsto \mathrm{Frm}'(\CO(\uparrow\! C\x\R),\ONS_{\uparrow C}),  \label{GTrhs}
\eeq
with the stipulation that
for $C\subseteq D$ the induced map  $$\mathrm{Frm}'(\CO(\uparrow\! C\x\R),\ONS_{\uparrow C})\raw  \mathrm{Frm}'(\CO(\uparrow\! D\x\R),\ONS_{\uparrow D})$$ is given by restricting an element of the left-hand side
to $\CO(\uparrow\! D\x\R)\subset \CO(\uparrow\! C\x\R)$; this is consistent by the same argument \er{con2}.

The Gelfand isomorphism \er{GTT} therefore becomes an arrow
\beq
\uA\sr{\cong}{\longrightarrow} \mathrm{Frm}'(\CO(\uparrow\! -\x\R),\ONS_{\uparrow -}),
\eeq
which, using \er{intA}, means that one has a compatible family of isomorphisms
\begin{eqnarray}
C & \sr{\cong}{\longrightarrow} &\mathrm{Frm}'(\CO(\uparrow\! C\x\R),\ONS_{\uparrow C});\nn\\
a &\mapsto& \hat{a}\inv: \CO(\uparrow\! C\x\R)\raw \ONS_{\uparrow C}. \label{GTfam}
\end{eqnarray}

On basic opens $\uparrow\! D \times U\in \CO(\uparrow\! C\x\R)$, with $D\supseteq C$, we obtain 
\begin{eqnarray}
\hat{a}\inv(\uparrow\! D \times U)&:& E\mapsto  [a\in U]\: \mathrm{if}\; E\supseteq D;
 \nn \\
& & E\mapsto 0 \: \mathrm{if}\; E\nsupseteq D, \label{GTtopos}
\end{eqnarray}
where $[a\in U]$ is the spectral projection of $a$ in $U$ (cf.\ \er{GTSP}); 
as it lies in $\CP(C)$ and  $C\subseteq D\subseteq E$,  $[a\in U]$ certainly lies in  $\CP(E)$, as required. 
We extend $\hat{a}\inv$ to general opens in $\uparrow\! C\x\R$ by the frame map property, and note that 
\er{phcon} for $\phv^*_C=\hat{a}\inv$ is satisfied.\footnote{It is not quite obvious that the Gelfand transform is an isomorphism, but this follows from the general theory \cite{BM}. In the special case that $C$ is maximal, though, in which case $\uparrow\!\! C=\{C\}$,
the isomorphism property is a consequence of the fact that any frame map $\CO(\R)\raw \CP(C)$ is of the form
$U\mapsto [a\in U]$ for some $a\in C\sa$ (there is an analogous statement for spectral measures). To prove this, note that $C\cong\C^n$, so that $\CP(\C)\cong \CO(\{1,2,\ldots,n\}$
(in the discrete topology).
Then because both $\{1,2,\ldots,n\}$ and $\R$ are sober, there is a bijection between frame maps
$\lm\inv :\CO(\R)\raw\CO(\{1,2,\ldots,n\})$ and (continuous) functions $\lm: \{1,2,\ldots,n\}\raw\R$. The $\lm(i)$ are the eigenvalues of $a$, and the given map  $\CO(\R)\raw \CP(C)$ provides the pertinent spectral projections.}
\section{Kochen--Specker Theorem}\setcounter{equation}{0}\label{s5}
 We now work out our abstract formulation \cite{HLS} of the Kochen--Specker Theorem \cite{KS}, with an indication how to prove it. Our reformulation is in the spirit of  Isham and Butterfield 
  \cite{BI2}, but we feel our version is more powerful, especially from a logical perspective.\footnote{Our statement that a certain locale has no points has a logical interpretation in terms of (the lack of) models of a so-called geometric theory \cite{HLS}, and is, in our opinion, considerably more revealing than the claim that some presheaf without further structure lacks global sections (i.e.\ points). As explained below, this logical perspective also suggests a wholly new type of proof of the Kochen--Specker Theorem.} Indeed, it is suggested by our formalism:
  \begin{quote}
{\it For $n>2$, the  Gelfand spectrum $\ulS$ of $\underline{M_n(\C)}$ has no points.}
\end{quote}
Some explanation is in order. We recall that the ``space'' $\ulS$ is really defined by its associated ``topology'', namely the frame $\OS$. Informally, a point of  $\ulS$ is an arrow
$\underline{\sg}: \underline{1}\raw\uSg$ in the topos $\TA$, but in the absence of $\ulS$ this formally denotes a frame map
\beq  \underline{\sg}^*:\OS\raw \CO(\underline{1})\equiv \uOm\label{pointint}
\eeq
 in $\TA$. Here $\uOm$ is the truth object or subobject classifier in $\TA$.
Hence our Kochen--Specker Theorem states that  there are no such frame maps. 

To show that this claim is equivalent to the usual Kochen--Specker Theorem, we use the external description of frames and frame maps (see Appendix \ref{A3} and the previous sections).
The external description of $\OS$ has already been given in \er{JTmap}; the external description of $\uOm$ is simply the identity map\footnote{For any space (or even locale) $X$, the external description of the
subobject classifier in $\Sh(X)$ is $ \mathrm{id}:\CO(X)\raw\CO(X)$.}
\beq \mathrm{id}: \CO(\CA)\raw \CO(\CA),\eeq
where $\CO(\CA)$ is the Alexandrov topology on the poset $\CA$.
Hence the  external description of \er{pointint} is 
 a frame map 
\beq
\sg^*: \ONS\raw \CO(\CA) \label{nonex}\eeq in \Sets\ 
 that satisfies the constraint
\beq \sg^*\circ\pi_{\Sg}^*=\mathrm{id}_{\CO(\CA)}. \label{constr}\eeq
If we regard \er{JTmap} as the inverse image map $\pi_{\Sg}^*=\pi_{\Sg}\inv$ of a ``virtual'' bundle \beq \pi_{\Sg}: \Sg\raw\CA,\label{vbundle}\eeq and similarly look at \er{nonex} as the inverse image map
$\sg^*=\sg\inv$ of a ``virtual'' continuous map $\sg:\CA\raw\Sg$, then the constraint \er{constr}
 is just the pullback of the rule $\pi_{\Sg}\circ\sg=\mathrm{id}_{\Sg}$
defining a continuous cross-section $\sg:\CA\raw\Sg$ of the virtual bundle \er{vbundle}.  In this virtual (or pointfree) sense, our Kochen--Specker Theorem therefore states that the bundle \er{vbundle} has no continuous cross-sections. 

Using \er{defpi}, we see that \er{constr} is explicitly given by
\beq \sg^*(\ch_{\uparrow C})=\uparrow\! C \label{C2}
\eeq
for all $C\in\CA$. We identify $\CO(\Sg)_{\uparrow C}$ in \er{bohrC} with
\begin{eqnarray}
\CO(\Sg)_{\uparrow C}&=& \{S\in\ONS\mid S(D)=0\:\forall\, D\nsupseteq C\}\nn \\
&=&  \{S\in\ONS\mid S\leq\ch_{\uparrow C} \}.
\end{eqnarray}
Since $\sg^*$ is a frame map, if $S\in \CO(\Sg)_{\uparrow C}$ and hence $S\leq\ch_{\uparrow C}$, then $\sg^*(S)\leq \sg^*(\ch_{\uparrow C})=\uparrow\! C$ by \er{C2}, so that $\sg^*$ restricts to a frame map \beq
\sg_C^*:  \CO(\Sg)_{\uparrow C}\raw \CO(\uparrow\! C).\label{ressg}
\eeq
Now take $a\in C\sa$. 
Combining \er{ressg} with the Gelfand transform \er{GTfam}, we obtain a frame map
\beq \sg_C^*\circ \hat{a}\inv: \CO(\uparrow\! C\x\R)\raw  \CO(\uparrow\! C).\label{combi}
\eeq
It can be shown that any frame map $\CO(\uparrow\! C\x\R)\raw  \CO(\uparrow\! C)$ is the inverse image map of a continuous function $\uparrow\! C\raw \uparrow\! C\x\R$,\footnote{
This is because $\CA$ and $\R$ are both sober; see footnote \ref{fnsober}. For $\R$ this is well known; we provide a proof for $\CA$. A space $X$  is sober iff each irreducible closed subset
 $S\subset X$ is the closure of a unique point in $S$. The closed subsets of a poset in the Alexandrov topology are the lower sets, and the closure of a point $x$ is the downset  $\downarrow\! x$. The  irreducible closed subsets are the directed  lower sets $S$ (i.e.\ 
 $S$ is a lower set and if $a,b\in S$ then there exists $c\in S$ such that $a\leq c$ and $b\leq c$
 (cf.\ \cite[p.\ 291]{SS}). We apply this to $X=\CA$. The directed sets $S$ in $\CA$ are those sets whose elements are mutually commuting subalgebras, and for which $C,D\in S$ implies the existence of $E\in S$ containing $C^*(C,D)$. Hence each directed lower set contains $C^*(C,D)$ whenever it contains $C$ and $D$. Since for $A=M_n(\C)$ each subset of $\CA$ that
 consists of mutually commuting subalgebras is finite, it follows that each  directed  lower set $S$ is the downset of the C*-algebra generated by the elements of $S$. This proves that $\CA$ is sober.
 }
 so that
\beq \sg_C^*\circ \hat{a}\inv=\til{V}_{(a,C)}\inv, \label{Cinv}
\eeq for some continuous $\til{V}_{(a,C)}:\uparrow\! C\raw \uparrow\! C\x\R$.
 Furthermore, the constraint \er{phcon} satisfied by $\phv_C^*=\hat{a}\inv$ and  the constraint \er{constr} satisfied by $\sg_C^*$ imply that $\til{V}_{(a,C)}$ takes the form 
 \beq 
 \til{V}_{(a,C)}(D)=(D,V_C(a)), \label{takes}\eeq
 for some $V_C(a)\in\R$. For fixed $C$ and $a\in C\sa$, 
 for each $D\supseteq C$ and $U\in\CO(\R)$ we define $S_{(D,U)}\in \CO(\Sg)_{\uparrow C}$ by
 \begin{eqnarray}
S_{(D,U)}(E) &=& [a\in U]\: \mathrm{if}\; E\supseteq D;
 \nn \\
& =& E\mapsto 0 \: \mathrm{if}\; E\nsupseteq D. \label{GTtopos8}
\end{eqnarray}
 Using \er{GTtopos} and \er{takes}, we then find that \er{Cinv} is equivalent to
 the following requirement:
 \begin{eqnarray}
\sg^*_C(S_{(D,U)})  &=& \uparrow\! D \: \mathrm{if}\; V_C(a)\in U;\nn \\
 &=& \emptyset\:  \mathrm{if}\; V_C(a)\notin U, \label{requ}
\end{eqnarray}
for all  $D\supseteq C$ and $U\in\CO(\R)$. Using the property
\beq [f(a)\in U]=[a\in f\inv(U)]\eeq
for each (bounded measurable) function $f:\R\raw\R$,
we infer that \er{requ} can only be consistent if
\beq V_C(f(a))=f(V_C(a)). \label{FUNC}\eeq
 Finally, take $D\supseteq C$. Because
$\sg^*_C$ and $\sg^*_D$ are both restrictions of the same map $\sg^*$ (see \er{ressg}), we have \beq
V_C(a)=V_D(a).\label{noncont}\eeq
Hence we may simply write $V_C(a)=V_D(a)=V(a)$, and regard the map $a\mapsto V(a)$ as a valuation that assigns a sharp value to each observable $a\in M_n(\C)\sa$. This valuation
satisfies the `functional composition principle' \er{FUNC} and the `noncontextuality requirement'
\er{noncont}, which are exactly the assumptions of the original Kochen--Specker Theorem \cite{KS} (see also \cite{MR} for a very clear discussion of these assumptions). We conclude that the statement that ``$\uSg$ has no points'' is equivalent to the  Kochen--Specker Theorem.

We close this section by giving yet another reformulation of it, which might have the advantage of admitting a direct proof (in our opinion,  the known proofs of  the  Kochen--Specker Theorem are obscure). 
Following the literature \cite{SS,MLM,J2}, we introduce\footnote{If $\Sg$ were a genuine space whose topology is sober, each frame map in \er{introduce} arises from a map $p:1\raw\Sg$, i.e.\ from a point of $\Sg$. Hence the name $\mathrm{pt}(\Sg)$.}
\beq \mathrm{pt}(\Sg)=\{\mbox{frame maps}\: p^*:\CO(\Sg)\raw\CO(1)\},\label{introduce} \eeq
simply defined in \Sets; recall that $1$ denotes any singleton set $1=\{*\}$. 
The set $\mathrm{pt}(\Sg)$ is topologized by declaring that the open sets are those of the form\beq
\mathrm{pt}(S)=\{p^*\in \mathrm{pt}(\Sg)\mid p^*(S)=*\},  \label{pt} \eeq
for each $S\in\CO(\Sg)$;
see \cite[\S IX.3]{MLM}. An alternative description of $\mathrm{pt}(\Sg)$ is as a subset 
$\mathrm{Pt}(\Sg)$ of $\ONS$, consisting of all $P\in \ONS$ that satisfy the conditions:
\begin{enumerate}
\item $P\neq \top$ (see \er{topS});
\item $U\wed V\leq P$ implies $U\leq P$ or $V\leq P$.
\end{enumerate}
The topology on $\mathrm{Pt}(\Sg)$ is given by the opens
\beq 
\mathrm{Pt}(S)=\{P\in \mathrm{Pt}(\Sg)\mid S\nleq P\}.\eeq
A homeomorphism $\mathrm{pt}(\Sg)\leftrightarrow  \mathrm{Pt}(\Sg)$, denoted by $p\leftrightarrow P$, is given by 
\begin{eqnarray}
P&=&\bigvee \{S\in\ONS\mid p^*(S)=\emptyset\}; \\
p^*(S) &=& \emptyset\:\:\: \mbox{ if }\: S\leq P;\\
 &=& *\:\:\:\mbox{ if }\: S\nleq P.
\end{eqnarray}

The point of introducing the space  $\mathrm{pt}(\Sg)$ is that any frame map $\sg^*:\CO(\Sg)\raw\CO(X)$ into the topology of a genuine space $X$ (which will be $\CA$ in what follows) factors as 
\beq \sg^*=\ovl{\sg}\inv\circ \mathrm{pt}, \label{factor} \eeq
for some continuous function $\ovl{\sg}: X\raw  \mathrm{pt}(\Sg)$, where $ \mathrm{pt}:\CO(\Sg)\raw \CO(\mathrm{pt}(\Sg))$
is given by \er{pt}. Conversely, any continuous $\ovl{\sg}$ defines a frame map $\sg^*$ by \er{factor} and hence one has a bijective correspondence between  frame maps $\sg^*:\CO(\Sg)\raw\CO(X)$ and continuous functions $\ovl{\sg}: X\raw  \mathrm{pt}(\Sg)$. Similarly with $\til{\sg}: X\raw  \mathrm{Pt}(\Sg)$.
Taking \er{constr} into account, our reformulation of the Kochen--Specker Theorem may then be expressed as follows:\begin{quote}
{\it There exists no continuous map $\ovl{\sg}:\CA\raw \mathrm{pt}(\Sg)$ that 
for each $C\in\CA$ satisfies $\ovl{\sg}\inv(\mathrm{pt}(\ch_{\uparrow C})=\uparrow\! C$. }
\end{quote}
Using $\mathrm{Pt}(\Sg)$ instead of $\mathrm{pt}(\Sg)$ (which has the advantage that continuity of $\til{\sg}:\CA\raw\mathrm{Pt}(\Sg)$  translates into inverse monotonicity),  the Kochen--Specker Theorem reads:
\begin{quote}
{\it There exists no map $\tilde{\sg}: \CA\raw \mathrm{Pt}(\Sg)$ that  for each inclusion $C\subseteq D$ satisfies  $\tilde{\sg}(D)\leq \tilde{\sg}(C)$ and for each pair $C,D\in\CA$ satisfies 
$\til{\sg}(D)\ngeq \ch_{\uparrow C}\: \Leftrightarrow\: C\subseteq D  $.
}
\end{quote}
\section{State-proposition pairing}\setcounter{equation}{0}\label{s7}
We now compute the particular state-proposition pairing \er{uom}, which we repeat for convenience:
\beq \big(\ueen\stackrel{\langle \ps,S\rangle}{\longrightarrow}\uOm\big)
\:=\: \big(\ueen \stackrel{S}{\raw} \CO(\uSg) \stackrel{\interpretation{\mu= 1}}{\longrightarrow}\uOm\big). \label{uom2}
\eeq
Here $\ps$ is a state on the \ca\ $M_n(\C)$, and $\mu$ is the induced probability measure (more precisely: valuation) on $\ulS$; see the Introduction.
As opposed to the usual probabilistic pairing taking values in the interval $[0,1]$, the
pairing \er{uom2} takes values in the subobject classifier $\uOm$ of $\TA$.
It can be shown (cf.\ Appendix \ref{A2}) that $\uOm(C)$ consists of all upper sets in $\uparrow\!\!C$, which means that each element $X\in \uOm(C)$ is a subset of $\CA$ with the properties that
$D\in X$ must satisfy $D\supseteq C$, and that if $D\in X$ and $E\supseteq D$, then $E\in X$. 
Each $\uOm(C)$ is a Heyting algebra in \Sets, partially ordered by set-theoretic inclusion, 
with the empty upper set $\emptyset$ as bottom element, and the maximal one $\uparrow\! C$ as top element. 

In principle, $\langle \ps,S\rangle$ is a natural transformation with components 
$\langle \ps,S\rangle(C)$ at each $C\in\CA$, but by naturality these are all given once the component at the bottom element $\C\cdot 1$ of $\CA$ is known. In somewhat sloppy notation, we may therefore
regard  $\langle \ps,S\rangle$, identified with $\langle \ps,S\rangle(\C\cdot 1)$, as an element of the set $\uOm(\C\cdot 1)$ of all upper sets in $\CA$. Consequently, $\langle \ps,S\rangle$ is simply a certain upper set in $\CA$, which turns out to be
\beq \langle \ps,S\rangle=\{C\in\CA\mid \ps({S}(C))=1\}.\label{result}\eeq
In words, the ``truth'' of $\langle \ps,S \rangle$ consists of those classical contexts $C$ in which the proposition $S(C)$ is true in the state $\ps$ in the usual sense (i.e.\ has probabilty one). 
Here we have identified the
arrow $S:\ueen\raw\OS$ in \er{uom2} with an element of $\ONS$ as given by \er{bohr} (see Section \ref{s3}),
so that $S(C)\in\CP(C)\subset M_n(\C)$, and hence $\ps({S}(C))$ is defined.\footnote{Expression  \er{result} would gain physical content when combined with the   D\"{o}ring--Isham `Daseinization' map \cite{DI,HLS}; see point 2 at the end of Section \ref{s3}. However, as it stands it already explicitly displays the non-binary truth values that are typical for the topos-theoretic approach to quantum physics.}

Let us note that \er{result} indeed defines an upper set in $\CA$. If $C\subseteq D$ then
${S}(C)\leq {S}(D)$, so that $\ps({S}(C))\leq \psi({S}(D))$ by positivity of 
states,\footnote{In case that $\ps$ is a vector state induced by a unit vector $\Psi\in\C^n$, this is the trivial property that $\Psi\in {S}(C)\C^n$ implies
$\Psi\in {S}(D)\C^n$.} so that $\ps({S}(D))=1$ whenever $\ps({S}(C))=1$
(given that $\ps({S}(D))\leq 1$, since $\ps(p)\leq 1$ for  any projection $p$).\footnote{In the vector state case, this means that  $C$ contributes to the upper set $\langle \ps,S\rangle$ iff $\Ps$ lies in the image of the
projection ${S}(C)$ in $\C^n$.  This is reminiscent of certain ideas in \cite{Isham}.}

To derive \er{result}, we note that $\mu:\CO(\uSg)\raw\ulR^+_l$ is a natural transformation,
defined by its components $\mu_C: \CO(\uSg)(C)\raw \ulR^+_l(C)$. It follows from
\cite[Corollary D4.7.3]{J2} that $\ulR^+_l(C)=L(\uparrow\!C,\R^+)$, the set of all lower semicontinuous functions from $\uparrow\!\!C$ to $\R^+$. Generalizing \er{mup} and making the identification \er {keyisoC}, it is easy to show that $$\mu_C:\CO(\uSg)(C)\raw L(\uparrow\!C,\R^+)$$ is given by
\beq \mu_C(S):\: \uparrow\!D\mapsto \ps({S}(D)),\eeq
for $D\supseteq C$ (see also Section 6 of \cite{HLS}). Finally, for $\sg:\CO(\uSg)\raw\ulR^+_l$ and  $\ta:\CO(\uSg)\raw\ulR^+_l$
one needs a formula for $\interpretation{\sg=\ta}_C:\CO(\uSg)(C)\raw\uOm(C)$, namely\footnote{Let $\mathbf{C}$ be a category with associated topos $\Sets^{\mathbf{C}^{\mathrm{op}}}$. Let $X,Y: \mathbf{C}^{\mathrm{op}}\raw\Sets$ be presheaves in this topos, let $\sg,\ta:X\raw Y$ and take $A\in \mathbf{C}$ and $x\in X(A)$. Then
the natural transformation $\interpretation{\sg=\ta}: X\raw\Om$ is given by its components
$\interpretation{\sg=\ta}_A: X(A)\raw\Om(A)$ as $x\mapsto\{f:B\raw A\mid \sg_B(Xf(x))=\ta_B(Xf(x))\}$.}
\beq \interpretation{\sg=\ta}_C(S)=\{D\in\uparrow\! C\mid \sg_D(S)= \ta_D(S)\}.\eeq
 \section{Parametrization of $\mathcal{C}(M_n(\C))$}\label{TOP}\setcounter{equation}{0}\label{s8}
 We start with a computation of the poset $\mathcal{C}(\Mt)$ of unital C*-subalgebras of \Mt. It is elementary that $\Mt$ has a single one-dimensional unital C*-subalgebra, namely $\C\cdot 1$, the multiples of the unit. Furthermore, any two-dimensional unital C*-subalgebra is generated by a pair of orthogonal one-dimensional projections.  
The one-dimensional projections in $M_2(\C)$ are of the form 
  \begin{equation}
p(x,y,z)=\frac{1}{2}   \left(\begin{array}{cc} 1+x & y+iz \\ y-iz & 1-x \end{array}\right),
\label{gens2} \end{equation}
where $(x,y,z)\in\R^3$ satisfies $x^2+y^2+z^2=1$. Thus the one-dimensional projections in $M_2(\C)$ are precisely parametrized by $S^2$. We obviously have $1-p(x,y,z)=p(-x,-y,-z)$, and since
the pairs $(p,1-p)$ and $(1-p,p)$ define the same C*-subalgebra, 
it follows that the two-dimensional elements of $\mathcal{C}(\Mt)$ may be parametrized by $S^2/\sim$, 
where $(x,y,z)\sim (-x,-y,-z)$ (in other words, antipodal points of $S^2$  are identified). This space, in turn, is homeomorphic with  the real projective plane $\mathbb{RP}^2$, i.e.\ the set of lines in $\R^3$ passing through the origin, or, equivalently, the space of great circles on $S^2$. 
This space has an interesting topology (which is quite different from the Alexandrov topology on the poset $\CA$), but in the present paper we ignore this aspect and just conclude that we may parametrize
\beq \mathcal{C}(\Mt)\cong *\cup \mathbb{RP}^2,\label{nis2}\eeq
where $*$ stands for  $\C\cdot 1$. A point $[x,y,z]\in  S^2/\sim$ then corresponds to the C*-algebra $C_{[x,y,z]}$ generated by the projections $\{p(x,y,z),p(-x,-y,-z)\}$. The poset structure of
$\mathcal{C}(\Mt)$ is evidently given by $*\leq [x,y,z]$ for any $[x,y,z]$ and no other relations.

 Let us now generalize the argument to determine $\mathcal{C}(M_n(\C))$ for any
$n$. In general, one has
\beq 
\mathcal{C}(M_n(\C))=\coprod_{k=1,\ldots, n} \mathcal{C}(k,n),\eeq
where  $\mathcal{C}(k,n)$ denotes the collection of 
all $k$-dimensional commutative unital C*-subalgebras of $M_n(\C)$.
To parametrize $\mathcal{C}(k,n)$, we note that each of its elements $C$ is a unitary rotation
$C=UDU^*$, where $U\in SU(n)$ and $D$ is some subalgebra contained in the algebra of all diagonal matrices. This follows from the case $k=n$, since each element of $\mathcal{C}(k,n)$ with $k<n$ is contained in some maximal abelian subalgebra.\footnote{For $k=n$, note that $C\in\CC(n,n)$ is generated by $n$ mutually orthogonal projections $p_1, \ldots, p_n$, each of rank 1.
Each  $p_i$ has a single unit eigenvector $u_i$ with eigenvalue 1; its other eigenvalues are 0. Put these $u_i$ as columns in a matrix, called $U$. Then $U^* p_i U$ is diagonal for all $i$: 
if $(e_i)$ is the standard basis of $\C^n$,  one has $Ue_i=u_i$ for all  $i$ and hence
$U^* p_i U e_i=U^* p_i u_i=U^* u_i=e_i$, while for $i\neq j$ one finds $U^* p_i U e_j=0$. Hence the matrix $U^* p_i U$ has a one at location $ii$ and zero's everywhere else.  All other elements $a\in C$ are functions of the $p_i$, so that $U^* a U$ is equally well diagonal. Hence $C=UD_nU^*$, with $D_n$ the algebra of all diagonal matrices.} Hence
\begin{equation}
\label{eq:Cnn}
  \mathcal{C}(n,n) = \{ U \cdot D_n  \cdot U^* \mid U \in SU(n)\},
  \end{equation}
  with $D_n=\{\mathrm{diag}(a_1,\ldots,a_n)\mid  a_i \in \field{C} \}$.
For $k<n$, $\mathcal{C}(k,n)$  is obtained by partitioning
$\{1,\ldots,n\}$ into $k$ nonempty parts, and
demanding $a_i=a_j$ for $i,j$ in the same part. However, because of
the conjugation with arbitrary $U \in SU(n)$ in \eqref{eq:Cnn}, two
such partitions induce the same subalgebra precisely when they permute
parts of equal size. Such permutations may be handled
using Young tableaux~\cite{F}. As the size of a part is of more
interest than the part itself, we define
\[
  Y(k,n) = \{ (i_1,\ldots,i_k) 
              \mid 0 < i_1 < i_2 < \cdots < i_k = n, \;\;
                   i_{j+1}-i_j \leq i_j-i_{j-1}
           \}
\]
(where $i_0=0$) 
as the set of partitions inducing different subalgebras. Hence
\begin{align*}
  \mathcal{C}(k,n) \cong \big\{ (p_1,\ldots,p_k) 
  & \;\colon p_j \in \mathcal{P}(\C^n), \;\; (i_1,\ldots,i_k) \in Y(k,n) \\
  & \mid \dim(\mathrm{Im}(p_j)) = i_j - i_{j-1}, 
    \;\; p_j \wedge p_{j'} = 0 \mbox{ for } j \neq j'
  \big\}.
\end{align*}
Now, since $d$-dimensional orthogonal projections in $\C^n$
bijectively correspond to the $d$-dimensional (closed) subspaces of
$\C^n$ they project onto, we can write
\[
  \mathcal{C}(k,n) \cong \big\{ (V_1,\ldots,V_k) 
  \;\colon (i_1,\ldots,i_k) \in Y(k,n),
    V_j \in \mathrm{Gr}(i_j-i_{j-1},n)
  \mid V_j \cap V_{j'} = 0 \mbox{ for } j \neq j'
  \big\},
\]
where \beq
\mathrm{Gr}(d,n) = U(n) / (U(d) \times U(n-d))\eeq is the well-known
Grassmannian, i.e.\ the set of all $d$-dimensional subspaces of
$\C^n$~\cite{GH}. In terms of the partial flag manifold
\beq
  \mathrm{G}(i_1,\ldots,i_k;n) 
  = \prod_{j=1}^k \mathrm{Gr}(i_j-i_{j-1}, n-i_{j-1}),
\eeq
 for $(i_1,\ldots,i_k) \in Y(k,n)$ (see~\cite{F}), we  finally obtain
\beq
  \mathcal{C}(k,n) \cong \{ V \in \mathrm{G}(i;n) \;\colon i \in
  Y(k,n) \} / \sim,
\eeq
where $i\sim i'$ if one arises from the other by permutations of 
equal-sized parts. 

Let us show how \er{nis2} is recovered from the above method. First, for any $n$ the set
 $\mathcal{C}(1,n)$ has a single element $*$, as there is only one
Young tableau for $k=1$. Second,  we have $Y(2,2)=\{(1,2)\}$, so
that 
\begin{align*}
      \mathcal{C}(2,2) 
  \cong \mathrm{G}(1,2;2) / S(2)
  = \mathrm{Gr}(1,2) \times \mathrm{Gr}(1,1) / S(2)
  \cong \mathrm{Gr}(1,2) / S(2)
  \cong \field{CP}^1 / S(2)
  \cong \field{RP}^2.
\end{align*}
\appendix
\section{Sheaves, topoi and Heyting algebras}\setcounter{equation}{0}
\subsection{Basic sheaf theory}\label{A1}
 We will only use a few basic categories:
\begin{enumerate}
\item The category $P$ defined by a partially ordered set (also called $P$), which is seen as a category by saying 
that $p,q\in P$ are connected by a single arrow iff $p\leq q$, with special cases:
\begin{enumerate}
\item  $P=\CO(X)$, the topology on a  space $X$, the partial order being by set-theoretic inclusion;
\item $P=\CA$, the poset of  unital commutative
C*-subalgebras of a unital \ca\ $A$ (which in this paper is usually $A=M_n(\C)$, the \ca\ of $n\x n$ complex matrices), again partially ordered by set-theoretic inclusion.
\end{enumerate}
\item The category \Sets\ of sets (as objects) and functions (as arrows), based on the usual ZF-axioms.
\item For any category $\CC$, the category $\widehat{\CC}=\Sets^{\CC^{\mathrm{op}}}$ of (covariant) functors
$\CC^{\mathrm{op}}\raw\Sets$, with natural transformations as arrows,\footnote{Here $\CC^{\mathrm{op}}$ is the opposite of a category $\CC$, which has the same objects as $\CC$ and also the same arrows, but the latter go in the opposite direction. If $\CC=P$ is a poset, this just means that in $P^{\mathrm{op}}$ the partial order is reversed.} with obvious special cases:
\begin{enumerate}
\item 
 $\TA=\Sets^{\CA}$;
\item  $\widehat{\CO(X)}=\Sets^{\CO(X)^{\mathrm{op}}}$, the category of so-called presheaves on a space $X$; 
\item The category $\Sh(X)$ of sheaves on $X$, which is the full subcategory of $\widehat{\CO(X)}$ defined by the following condition: a presheaf $F:\CO(X)^{\mathrm{op}}\raw\Sets$ is a sheaf if for any open $U\in\CO(X)$, any open cover $U=\cup_i U_i$ of $U$, and any family $\{s_i\in F(U_i)\}$ such that $F(U_{ij}\leq U_i)(s_i)=F(U_{ij}\leq U_j)(s_j)$ for all $i,j$, there is a unique $s\in F(U)$ such that $s_i=F(U_i\leq U)(s)$ for all $i$. Here $U_{ij}=U_i\cap U_j$ and $F(V\leq W):F(W)\raw F(V)$ is the arrow part of the functor $F$, defined whenever $V\subseteq W$.
Using the concept of a limit, we may write this as 
\beq F(U) = \underleftarrow{\lim}_i F(U_i). \label{limit}\eeq
\end{enumerate}
\hyphenation{ho-meo-mor-phism}
 \item The category $E(X)$ of \'{e}tale bundles $\pi:B\raw X$ over $X$, where $\pi$ is a local homeomorphism,\footnote{That is, each $p\in B$ has an open neighbourhood $U$ for which $\pi(U)$ is open in $X$ and homeomorphic to $U$ through $\pi$ \cite[\S II.6]{MLM}.}  and the arrows between
$\pi_Y:Y\raw X$ and $\pi_Z:Z\raw X$ are those continuous maps $\ps: Y\raw Z$ that satisfy $\pi_Z\circ\ps=\pi_Y$.
\end{enumerate}

It is important  for some of the more technical arguments in this paper that for any poset $P$, the category $\Sets^P$ 
is equivalent to the category $\Sh(P)$ of sheaves on $P$ with respect to the so-called  Alexandrov topology (see footnote \ref{AT}).
 The equivalence
 \begin{equation}\label{Al}
  \Sets^P\simeq \Sh(P)
\end{equation}
is given by mapping a functor  $\uF:P\raw\Sets$ to a sheaf  $F:\CO(P)^{\mathrm{op}}\raw\Sets$ by 
defining the latter on basic opens by
\beq   F(\uparrow\! x) = \uF(x), \label{FFtil} \eeq
extended to general Alexandrov opens by \er{limit}. 
Vice versa,  a sheaf $F$  on $P$ immediately defines  $\uF$ by reading \er{FFtil} from right to left. In particular, we have
\beq   
\asstopos(\alg{A})\simeq \Sh(\CA),\label{TASh} 
\eeq
where $\CA$ is understood to be equipped with the Alexandrov topology.

Another useful equivalence is $E(X)\simeq \Sh(X)$. The functor $E(X)\raw\Sh(X)$ establishing half of this equivalence
is given by $\til{F}\mapsto F$, $F(U)=\Gm(U,\til{F})$, that is, the set of continuous cross-sections $U\raw\til{F}$, with functoriality given by restriction. In the opposite direction, the pertinent functor $\Sh(X)\raw E(X)$ associates a bundle $\til{F}$ to a sheaf $F$ whose fiber (or ``stalk'') at $x\in X$ is  
\beq\til{F}_x=\underrightarrow{\lim}_{\CO_x(X)}F(U)\cong \{[s]_x\mid s\in F(U), U\in \CO_x(X)\}.\eeq
Here the colimit  in the first expression is taken over all $U\in\CO_x(X)$, the set  of all open neighbourhoods of $x$. The right-hand side provides  an explicit expression for this colimit, in which the equivalence class $[s]_x$ (the germ of $s$ at $x$) is defined by saying that $s\sim_x t$ for $s\in F(U)$ and $t\in F(V)$, $U,V\in \CO_x(X)$, when there exists $W\in \CO_x(X)$ such that $W\subseteq U\cap V$ and  $s_{|W}=t_{|W}$. The topology on $\til{F}$ is given by declaring that \beq
\mathcal{B}(\til{F}):=\{\dot{s}(U)\mid U\in\CO(X), s\in F(U)\}\label{topB} \eeq
is a basis of $\CO(\til{F})$, where the cross-section $\dot{s}:U\raw\til{F}$ is defined by
$\dot{s}(x)=[s]_x$. See \cite[\S II.6]{MLM}. 
This equivalence assumes a particularly simple form when $X=P$ is a poset equipped with the Alexandrov topology \cite[\S 14.1]{Gold}. In that case, the bundle
defined by a functor $\uF:P\raw\Sets$, or rather by the associated sheaf $F$ as in \er{FFtil}, has fibers
\beq \til{F}_x=\uF(x)\label{fibers} \eeq and a topology generated by the basis 
\begin{eqnarray}
 \mathcal{B}(\til{F})& =& \{B_{x,s}\mid x\in P,\, s\in \uF(x)\}; \nn \\  
B_{x,s}& =& \{\uF(x\leq y)(s)\mid y\in P,\, x\leq y\},
\label{topology}
\end{eqnarray}
where $\uF(x\leq y):\uF(x)\raw \uF(y)$ is the arrow part of $\uF$.
This follows from \er{topB} because for $U,V\in \CO_x(X)$ there is a smallest $W\subseteq U\cap V$ containing $x$, namely $\uparrow\!x$. Consequently, if $U=\uparrow\!x$ and $s\in \Gm(U,\til{F})$ in the above analysis, one has $\dot{s}=\uF(x\leq y)(s)$, which leads to \er{topology}.

For example, our internal \ca\ $\uA$ may be described as an \'{e}tale bundle $\til{A}$. It is immediate from \er{intA} and \er{fibers} that the fibers of $\til{A}$ are
\beq \til{A}_C=C, \label{AasB}\eeq
so that in passing from $A\in\TA$ to $\til{A}\in E(\CA)$ we have replaced the tautological functor by the tautological bundle. 
According to \er{topology}, the topology on $\til{A}$ is generated by the basis opens
\beq B_{a,C}=\{a\in D\mid D\supseteq C\}.\label{aC}\eeq
Here $a\in C$, and the open \er{aC} tracks this element as it embeds in all possible $D$'s in which $C$ is contained; here $a\in D$ is an element of the fiber $\til{A}_D$ above $D$, to be distinguished from $a\in C$ which lies in the fiber $\til{A}_C$ above $C$.

\subsection{Basic topos theory}\label{A2}
This paper is mainly concerned with the categories $\TA$ and $\Sh(\CA)$, the latter with respect to  the Alexandrov topology. 
These categories are examples of topoi.
 The advantage of working in a topos is that most set-theoretic reasoning can be carried out,
 with the restriction that all proofs need to be constructive (i.e.\ cannot make use of the law of the excluded middle or the Axiom of Choice).
Specifically,
a topos is a category with the following ingredients (all unique up to
  isomorphism):
\begin{enumerate}
\item {\it Terminal object.} This is an object called 1 such that for each object $A$ there is a unique arrow
  $A\raw 1$. In \Sets\ this is any  singleton set $*$. In $\TA$ and $\Sh(\CA)$ (and more generally in $\widehat{\CC}$),
   it is the constant functor taking the value $*$.
\item {\it Pullbacks.} These generalize the fibered product $B\x_A
  C=\{(b,c)\in B\x C\mid f(b)=g(c)\}$  of $B\sr{f}{\raw}A$ and
  $C\sr{g}{\raw}A$ in \Sets\ into a pullback square with appropriate
  universality property.  Cartesian products are a special case. In $\widehat{\CC}$ (and hence in $\TA$ and $\Sh(\CA)$), pullbacks may be computed ``pointwise''; see \cite[\S I.3]{MLM}.
\item {\it Exponentials.} These generalize the idea that the class
  $B^A$ of functions from a set $A$ to a set $B$ is itself a set, and
  hence an object in \Sets, equipped with the evaluation map
  $\mathrm{ev}:A\x B^A\raw B$. In $\Sh(X)$ one may take \cite[\S II.8]{MLM}
  \beq F^G(U)=\mathrm{Nat}(G_{|U},F_{|U}), \label{NatU}\eeq
  the set of natural transformations between the functors $G$ and $F$, both restricted to $\CO(U)$
  (i.e.\ defined on each open $V\subseteq U$ instead of all $V\in\CO(X)$). The map $\mathrm{ev}_U:
  G(U)\x \mathrm{Nat}(G_{|U},F_{|U})\raw F(U)$ is the obvious one, sending $(g,\theta)$ to $\theta_U(g)$.
  By \er{TASh}, in $\TA$ one analogously has
    \beq \uF^{\underline{G}}(C)=\mathrm{Nat}(\underline{G}_{\uparrow C},\uF_{\uparrow C}) \label{NatC}\eeq
  at each $C\in\CA$, where $\uF_{\uparrow C}$ is the restriction of the functor $\uF:\CA\raw\Sets$ to
  $\uparrow C\subseteq\CA$. In particular, since $\C\cdot 1$ is the bottom element of the poset $\CA$, one has
  \beq  \uF^{\underline{G}}(\C\cdot 1)=\mathrm{Nat}(\underline{G},\uF).\label{iP}\eeq
\item {\it Subobject classifier.} This generalizes the idea that one
  may characterize a subset $A\subseteq B$ by its characteristic
  function $\ch_A:B\raw \{0,1\}$. Subsets generalize to subobjects,
  i.e.\  monic (``injective") arrows $A \rightarrowtail B$, and in a
  topos there exists an object $\Om$ (the subobject classifier) with
  associated arrow $1\sr{\top}{\raw}\Om$ (``truth'') such that for any
  subobject $A \rightarrowtail B$ there is a {\it unique} arrow
  $B\stackrel{\chi_A}{\longrightarrow}\Om$ for which
  $B\sr{f}{\law}A\raw 1$ is a pullback of
  $B\stackrel{\chi_A}{\longrightarrow}\Om$ and
  $1\sr{\top}{\raw}\Omega$. Conversely, given any arrow
  $B\sr{\ch}{\raw}\Om$ there exists a subobject $A \rightarrowtail B$
  of $B$ (unique up to isomorphism) whose classifying arrow $\ch_B$
  equals $\ch$. The subobject classifier in a topos play the role of a ``multi-valued truth object'', generalizing the simple situation in \Sets, where  $\Om=\{0,1\}=\{\mathrm{false, true}\})$.
  
 In $\Sh(X)$ the subobject classifier is the sheaf $\Omega: U\mapsto  \CO(U)$ with $\Omega(V\leq U):\CO(U)\raw \CO(V)$
 given by $W\mapsto W\cap V$ whenever $V\subseteq U$; see \cite[\S II.8]{MLM}. The truth map $1\sr{\top}{\raw}\Om$
 is given at $U$ by $\top_U(*)=U$.
 Hence in our topos $\TA$  the subobject classifier  $\functor{\Omega}$ is the functor assigning to $C \in
\CA$ the collection $UC$ of upper sets on $C$,\footnote{This means that $X\subset \CA$ lies in $UC$ if $X\subseteq \uparrow\! C$ and if
$D\in X$ and $D\subset E$ implies
$E\in X$.} and to an arrow $C\subseteq D$ in $\CA$ the obvious map $UC\raw UD$ given by $X\mapsto X\cap \uparrow\! D$.
  \end{enumerate}
  
Combining the third and fourth points,  one has 
 \beq \uOm^{\uF}(C)\cong\mathrm{Sub}(\uF_{\uparrow C}),
 \label{Yoneda}
 \eeq  the set of subfunctors of $\uF_{\uparrow C}$. 
  In particular, like in \er{iP} we find
 \beq\Om^{\uF}(\C\cdot 1)\cong \Hom({\uF},\Om)\cong \mathrm{Sub}({\uF}),\eeq the set of subfunctors of ${\uF}$ itself. 
 If $C\subseteq D$, then the map $\Om^{\uF}(C)\raw \Om^{\uF}(D)$  defined by $\Om^{\uF}$, identified with a map $\mathrm{Sub}({\uF}_{\uparrow C})\raw \mathrm{Sub}({\uF}_{\uparrow D})$, is simply given by restricting a given subfunctor of ${\uF}_{\uparrow C}$
 to $\uparrow\!D$. 
 \subsection{Heyting algebras and frames}\label{A3}
A  {\it Heyting algebra}
is a distributive lattice $\CL$ with a map $\raw:\CL\x\CL\raw\CL$ (called implication)
satisfying 
\beq
x\leq (y\raw z)\:\:\mbox{iff}\:\: x\wed y\leq z.\eeq  Every Boolean algebra is
a Heyting algebra, but not {\it vice versa}; in fact, a Heyting
algebra is Boolean iff $\neg\neg x=x$ for all $x$, which is the case
iff $\neg x\vee x=\top$  for all $x$. Here negation is a derived
notion, defined by 
\beq \neg x=(x\raw\perp). \label{defneg}
\eeq
A  Heyting algebra is {\it complete} when arbitrary
joins (i.e.\  sups) and meets (i.e.\  infs) exist.

A {\it frame} is  a complete distributive lattice such that
$x\wedge \bigvee_{\lambda}y_{\lambda}=\bigvee_{\lambda}x\wedge
y_{\lambda}$ for arbitrary families $\{y_{\lambda}\}$ (and not just for finite ones, in which case the said property follows from the definition of a distributive lattice).  For  example, if $X$ is a topological space, then the
topology $\CO(X)$ of $X$ is a frame with $U\leq V$ if $U\subseteq V$.
A {\it frame map} preserves finite meets and arbitrary
joins; this leads to the category $\mathbf{Frm}$  of frames and frame maps. 
For example, if $f:X\raw Y$ is continuous then $f\inv: \CO(X)\raw \CO(Y)$ is a frame map. For this reason, frames are often  denoted by $\CO(X)$ and frame maps 
are written $f\inv$ or $f^*$, even if the frame does not come from a topological space. 

Any frame is at the same time a complete Heyting algebra, with implication
\beq
x\raw y=\bigvee\{z\mid z\wed x\leq y\}. \label{fHA}
\eeq In particular, it follows from \er{defneg} and \er{fHA} that
\beq \neg x=\bigvee\{z\mid z\wed x=\bot\}.\label{negx}\eeq
 Conversely, the infinite distributivity law in a frame is
automatically satisfied in a Heyting algebra, so that frames and complete 
Heyting algebras are essentially the same things.\footnote{They do not form isomorphic or even equivalent categories, though, for frame maps do not necessarily preserve the implication $\raw$ defining the Heyting algebra structure.}

Frames can be defined internally in any topos, and those in $\Sh(X)$ can be described explicitly \cite{Joh,JT} (see also \cite[\S C1.6]{J2}).
Namely, there is an equivalence  
\beq \mathbf{Frm}_{\Sh(X)}\simeq(\mathbf{Frm}_{\Sets}/\CO(X))^{\mathrm{op}} \label{JTE} \eeq
between the category  of internal frames in  $\Sh(X)$ and the category of
frame maps in \Sets\ with domain $\CO(X)$, where the arrows between two such maps
\begin{eqnarray}
\pi_Y^*&:&\CO(X)\raw\CO(Y);\label{fm1}\\
\pi_Z^*&:&\CO(X)\raw\CO(Z)\label{fm2};
\end{eqnarray}
are the frame maps 
\beq
\phv^*:\CO(Z)\raw\CO(Y) \label{fm3}
\eeq satisfying\footnote{This looks more palpable in terms of the ``virtual'' underlying spaces. If 
\er{fm1} - \er{fm3} are seen as inverse images $\pi^*=\pi\inv$ of maps
$\pi_Y:Y\raw X$ , $\pi_Z:Z\raw X$ and $\phv: Y\raw Z$, then \er{fm4} corresponds to $\pi_Z\circ\phv=\pi_Y$.}
\beq
\phv^*\circ\pi_Z^*=\pi_Y^*.\label{fm4}\eeq
In this paper, this characterization is used to compute the frame maps in 
$\mathbf{Frm}_{\Sh(X)}$, whose internal characterization is rather indirect. 

The equivalence \er{JTE} comes about as follows. First, a frame map $\pi_Y^*:\CO(X)\raw\CO(Y)$ defines an internal frame
$ \CO(\CI_{\pi})$ in the topos $\Sh(X)$ as the sheaf 
\beq \CO(\CI_{\pi}):U\mapsto\,\downarrow\!\pi^*(U)\equiv
 \{W\in\CO(Y)\mid W\leq \pi^*(U)\},\label{Ipi}
\eeq
with $\CO(\CI_{\pi})(U\leq V):\: \downarrow\!\pi^*(V)\raw \downarrow\!\pi^*(U)$ given by intersection with
$\pi^*(U)$.
Given  frame maps \er{fm1} -\er{fm3}, one obtains an internal frame map $\underline{\phv}^*:\CO(\CI_{\pi_Z})\raw\CO(\CI_{\pi_Y})$ in $\Sh(X)$ by defining its components as a natural transformation by
\begin{eqnarray}
\underline{\phv}^*(U):\: \downarrow\!\pi_Z^*(U)&\raw& \downarrow\!\pi_Y^*(U);\nn \\
S &\mapsto& \phv^*(S).\label{simplyput}
\end{eqnarray}
Conversely, let  $\CO(\uSg)$ be an internal frame in $\Sh(X)$. 
Consider 
\beq \ONS=\CO(\uSg)(X),\eeq which is a frame in $\Sets$. Define a map
\beq
\pi_{\Sg}^*:\CO(X)\raw \CO(\Sigma)\label{pistar}
\eeq by
\beq \pi_{\Sg}^*(U)=\bigwedge\{S\in \CO(\Sigma)\mid \CO(\uSg)(U\leq X)(S)=\top\},
\label{keyfor}\eeq
 where $\top$ is the top element of the complete lattice $\CO(\uSg)(U)$, and the map  $$\CO(\uSg)(U\leq X):\CO(\uSg)(X)\raw \CO(\uSg)(U)$$ is defined by the arrow part of the functor $\CO(\uSg):\CO(X)^{\mathrm{op}}\raw\Sets$. 
Then \er{pistar} is a frame map, whose corresponding internal frame $\CI_{ \pi_{\Sg}}$ is isomorphic to $\CO(\uSg)$. 
The map \er{pistar} is called the {\it external} description of $\CO(\uSg)$.\footnote{An important application is the external reformulation of internal properties of $\CO(\uSg)$ in terms of set-theoretic properties of the map \er{pistar}. For example, the general theory of \cite{BM} requires that the Gelfand spectrum $\OS$ of our internal \ca\ $\uA$ has a technical property called {\it regularity} (which is a frame-theoretic generalization of the well-known corresponding property for topological spaces)
\cite{SS}. This internal property may indeed be verified from the external version of regularity given in \cite{J3}.} 

In order to determine specific frames in $\Sh(X)$, we need a further result from topos theory. The Dedekind real numbers $\R_d$ and the lower real numbers $\R_l$ (which
describe sets of the type $x<q$, $q\in\Q$) can both be axiomatized by what is called a
geometric propositional theory $\T$.  In any topos $\CT$ (with so-called natural numbers object), such a theory determines a certain frame $\CO(\T)_{\CT}$, whose ``points'' are defined as frame maps $\CO(\T)\raw\Om$, where $\Om$ is the subobject classifier in $\CT$ (more precisely, the object of points of $\CO(\T)$ in $\CT$ is the subobject of
$\Om^{\CO(\T)}$ consisting of frame maps).
 For example, if $\T_{\R_d}$ is the theory axiomatizing $\R_d$, in \Sets\ one simply
 has 
 \beq \CO(\T_{\R_d})_{\Sets}=\CO(\R),\label{Rdtop}
 \eeq whose points comprise the set $\R$. 

The key result is as follows. Let $\pi_{\T}:X\x \CO(\T)_{\Sets}\raw X$ be projection on
the first component, with associated frame map $\pi_{\T}^*\equiv\pi_{\T}\inv:\CO(X)\raw \CO(X\x \T_{\Sets})$. Then
\beq \CO(\T)_{\Sh(X)}=\CO(\CI_{\pi_{\T}}). \label{intsh}\eeq
Using \er{Rdtop}, this yields that the frame of Dedekind real numbers
 $\CO(\R_d)\equiv\CO(\T_{\R_d})$ is the sheaf
\beq \CO(\R_d)_{\Sh(X)}: U\mapsto \CO(U\x \R),\eeq
whereas the Dedekind real numbers object is the sheaf
\beq (\R_d)_{\Sh(X)}: U\mapsto C(U,\R).\label{DRN}\eeq
Similarly, for the lower real numbers (whose frame we will not need) one obtains 
\beq (\R_l)_{\Sh(X)}: U\mapsto L(U,\R),\label{LRN}\eeq
 the set of all lower semicontinuous functions from $U$ to $\R$ that are locally bounded from above  \cite[Corollary D4.7.3]{J2}. Using \er{FFtil}, such results may immediately 
be transferred to $\Sets^P$ and hence to $\TA$. 
For example, one has
\beq \CO(\underline{\R}_d)\equiv \CO(\R_d)_{\TA}: C\mapsto \CO((\uparrow\! C)\x \R). \label{ddr}\eeq
Since Alexandrov-continuous functions must be locally constant, it follows from \er{DRN} that
\beq \underline{\R}_d: C\mapsto\R.\eeq
For the lower reals, however, \er{LRN} yields
\beq \underline{\R}_l: C\mapsto L(\uparrow\!C,\R).\eeq

 \end{document}